\documentclass[aps,prf,reprint,onecolumn,groupedaddress]{revtex4-1}

\usepackage{amsmath}

\usepackage{multirow}

\usepackage{siunitx}
\usepackage{gensymb}  

\usepackage{booktabs}

\usepackage{pgfplots}  
\pgfplotsset{compat=newest}
\usepackage{tikz}
\usetikzlibrary{plotmarks,backgrounds,patterns,shapes,external,pgfplots.groupplots}
\tikzset{external/only named=true}

\tikzset{cross/.style={cross out, draw=black, minimum size=2*(#1-\pgflinewidth), inner sep=0pt, outer sep=0pt},
cross/.default={1pt}}
\newcommand\spiral{}%
\def\spiral[#1](#2)(#3:#4:#5){%
\pgfmathsetmacro{\domain}{pi*#3/180+#4*2*pi}
\draw [#1,shift={(#2)}, domain=0:\domain,variable=\t,smooth,samples=int(\domain/0.08)] plot ({\t r}: {#5*\t/\domain})
}

\usepackage{hyperref}
\usepackage[noabbrev,capitalize]{cleveref}

\DeclareMathOperator{\dx}{\,dx}

\DeclareMathOperator{\Rey}{\mbox{\textit{Re}}}
\newcommand{\Bo}{\operatorname{\mbox{\textit{Bo}}}}

\newcommand{\Ca}{\operatorname{\mbox{\textit{Ca}}}}
\newcommand{\A}{\operatorname{\mbox{\textit{A}}}}
\newcommand{\Cn}{\operatorname{\mbox{\textit{Cn}}}}
\newcommand{\Pe}{\operatorname{\mbox{\textit{Pe}}}}
\newcommand{\Co}{\operatorname{\mbox{\textit{Co}}}}

\begin{document}

\title{On the stability of gravity-driven liquid films overflowing microstructures with sharp corners}

\author{Henning Bonart}
\email[]{henning.bonart@tu-berlin.de}
\affiliation{Technische Universit\"at Berlin, Process Dynamics and Operations Group,\\Straße des 17. Juni 135, 10623 Berlin, Germany}

\author{Johannes Jung}
\affiliation{Technische Universit\"at Berlin, Process Dynamics and Operations Group,\\Straße des 17. Juni 135, 10623 Berlin, Germany}

\author{Jens-Uwe Repke}
\email[]{jens-uwe.repke@tu-berlin.de}
\homepage[]{www.dbta.tu-berlin.de}
\affiliation{Technische Universit\"at Berlin, Process Dynamics and Operations Group,\\Straße des 17. Juni 135, 10623 Berlin, Germany}

\date{\today}

\begin{abstract}
We are concerned with the stability of thin liquid films overflowing single microstructures with sharp corners.
The microstructures were of rectangular and triangular shape.
Their heights and widths were $0.25$, $0.5$ and $0.75$ times the Nusselt film thickness.
To observe smooth, wavy and very unstable films we performed simulations with Reynolds numbers ranging from 10 to 70.
The dynamics of the liquid film and the overflowing gas phase were described by the coupling between the Cahn--Hilliard and Navier--Stokes equations. 
The resulting model forms a very tightly coupled and nonlinear system of equations.
Therefore we carefully selected the solution strategy to enable efficient and accurate large-scale simulations.
Our results showed that the formation of waves was shifted to higher Reynolds numbers compared to the film on a smooth surface. 
If waves were finally formed the microstructures led to irregular waves.
Our results indicate a great influence of the microstructure's shape and dimension on the stability of the overflowing liquid film.

\end{abstract}


\maketitle
 
\section{Introduction}
Gravity-driven liquid films overflowing a solid, structured surface appear in numerous technical applications.
Examples include falling film evaporators or structured high performance packing for absorption columns.
For research conducted on the stability of thin liquid films flowing on smooth and flat substrates and the formation of waves we refer to the recent review by~\citet[Ch.5]{Aksel2018}.
In the literature, some results were documented on the influence of smooth microstructures, like hemispheres (\cite{Blyth2006, Veremieiev2015}), on the stability of liquid films, see the review by~\citet[Ch. VII. A]{Craster2009}.
It is well-known for single phase flows, that sharp corners have a great influence on stability and flow separation, see for example~\cite{Brunold1989,Ozgoren2006,Hu2006,Alam2011}.
Surprisingly, to the best of our knowledge only few papers discuss results on thin liquid films and single microstructures with sharp corners and positive elevation in the size of the film itself.
Almost sharp structures like rectangles with rounded corners were simulated by~\cite{Tseluiko2013}.
Results on trenches or single step-ups or -downs with sharp corners were described by~\cite{Gaskell2004, Veremieiev2010, Veremieiev2015}.
However, despite their relevance in technical applications, systematic studies of the influence of single, sharp microstructures on the stability of the film are rarely found.

In this paper we report on detailed numerical simulations of films overflowing single microstructures with sharp corners.
The dynamics of the two-phase flow are described by the coupling between the Cahn--Hilliard (CH) and the Navier--Stokes (NS) equations. 
In this way, we are able to fully resolve the sharp corners.
The resulting model forms a very tightly coupled and nonlinear system of equations.
Therefore we carefully select the solution strategy to enable efficient and accurate simulations.
This includes the linearization and decoupling of the equations and preconditioned Krylov methods for the solution of the arising linear systems.
The examined microstructures are of rectangular and triangular shape with heights and widths of 0.25, 0.5 and 0.75 compared to the particular Nusselt film thickness.
The Reynolds numbers range from 10 to 70.

The remainder of the article is structured as follows: 
In~\cref{sec:chns}, we discuss the governing equations and present the system in nondimensional form.
The numerical method is described in~\cref{sec:numerics}. 
Here, we give details on the decoupling of the equations, the descretization and the preconditioner.
The test case is described in~\cref{sec:mesh}.
Finally, we present and discuss our results in~\cref{sec:results}.
In~\cref{sec:conclusion} we conclude our paper.

\section{Cahn--Hilliard--Navier--Stokes Equations}\label{sec:chns}
We treated the liquid film as well as the overflowing gas phase as Newtonian, isotherm, immiscible and incompressible fluids.
The thermodynamically consistent Cahn--Hilliard--Navier--Stokes model presented in~\cite{Abels2012} was applied.
The model combines the common incompressible, single-field Navier--Stokes (NS) equations with the convective Cahn--Hilliard (CH) equations to describe the interface dynamics between the liquid and the gas.
It is given by:
\begin{align}
\rho\partial_t v + ((\rho v + J)\cdot\nabla) v -\mbox{div}\left(2\eta Dv\right) + \nabla p &= -\varphi\nabla \mu + \rho g\;, \label{eqn:ns1}\\
-\mbox{div}(v) &= 0\label{eqn:ns2}\;,\\
\partial_t \varphi + v \cdot\nabla \varphi - b\Delta \mu &= 0\;,\label{eqn:ch1}\\
-\sigma\epsilon\Delta \varphi + \frac{\sigma}{\epsilon}W^\prime(\varphi) &= \mu\;,\label{eqn:ch2}
\end{align}
with the velocity field $v$, the pressure field $p$, the phase field $\varphi$ and the chemical potential $\mu$.
The velocity deformation tensor and the gravitational acceleration are given by $2Dv := \nabla v + (\nabla v)^t$ and $g$.
The density function is denoted by $\rho(\varphi)$ and satisfies $\rho(-1) = \rho_1$ and $\rho(1) = \rho_2$, with $\rho_2 > \rho_1 > 0$ denoting the constant densities of the two involved fluids. 
The viscosity function is $\eta(\varphi)$ and satisfies $\eta(-1) = \eta_1$ and $\eta(1) = \eta_2$, with $\eta_1,\eta_2$ denoting the viscosities of the involved fluids.
For this work, they were chosen as:
\begin{align}
		\rho = \frac{1}{2}\left((\rho_l + \rho_g) + \varphi (\rho_l - \rho_g)\right)\;,\;\;\;
		\eta = \frac{1}{2}\left((\eta_l + \eta_g) + \varphi (\eta_l - \eta_g)\right)\;.
\end{align}

\Cref{eqn:ns1,eqn:ns2} are the common incompressible Navier--Stokes equations with two additional terms: 
The density flux $J := -b\frac{\partial\rho}{\partial\varphi}\nabla \mu$ guarantees consistency and enhances stability if the densities of the two fluids are different.
The surface tension force is modelled by $\varphi\nabla\mu$.
The parameter $b$ stems from the diffuse interface approach in the Cahn--Hilliard equation~\cref{eqn:ch1,eqn:ch2}.
It represents the mobility of the two-phase interface.
The thickness of the diffuse interface is described by $\epsilon$.
The scaled surface tension is given by $\sigma = \frac{3}{2\sqrt{2}}\sigma^{phy}$ with the physical surface tension $\sigma^{phy}$.
The function $W(\varphi)$ denotes a dimensionless potential of double-well type with two strict minima at $\pm 1$.
Here, we chose it as
\begin{align}
W(\varphi) := 
	\begin{cases}
		\frac{1}{4}(1-\varphi^2)^2 & \mbox{if } |\varphi|\leq 1,\\
		(|\varphi|-1)^2 & \mbox{else.}
	\end{cases} 
\end{align}
For different choices of $W$ and a comparison we refer to~\cite{Bonart2019b}.

The model \eqref{eqn:ns1}--\eqref{eqn:ns2} can be derived purely from thermodynamic principles~\cite{Abels2012}.
It is postulated that the system in the whole domain can be described by the following sum of kinetic energy and Helmholtz free energy functional of Ginzburg--Landau type~\cite{Jacqmin1999}
\begin{align}
		E = \frac{1}{2}\int_\Omega \rho |v|^2\;\dx + \sigma\int_\Omega \epsilon^{-1}W(\varphi) + \epsilon|\nabla \varphi|^2\;\dx.
\end{align}
Compared to sharp interface methods, the phase field method replaces the infinitely thin boundary between gas and liquid by a transition region with finite thickness.
It describes the distribution of the different fluids by a smooth indicator function where -1 is pure gas and +1 is pure liquid.
It follows, that all physical properties like density or viscosity vary continuously across the interface. 
As summarized in the review by~\cite{Worner2012}, the Cahn--Hilliard--Navier--Stokes (CHNS) equations can easily handle large topological changes of the interface~\cite{Anderson1998} and the interface is implicitly tracked without any prior knowledge of the position.
Furthermore, one of the major advantages is that the formulation of the surface tension force in the Navier--Stokes (NS) equation conserves both the surface tension energy and kinetic energy. 
This can reduce spurious currents, which are purely artificial velocities around the interface, to the level of the truncation error even for low Capillary numbers~\cite{He2008,Jamshidi2018}.

\subsection{Nondimensionalization}\label{sec:nondim}
We scale the coupled CHNS system~\cref{eqn:ns1,eqn:ns2,eqn:ch1,eqn:ch2} with:
\begin{align}
t = \frac{\hat t L}{U}\;,\;\;
x = L \hat x\;,\;\;
v = U \hat v\;,\;\;
p = U^2 \rho_l\hat p\;,\;\; %
\mu = \frac{U \eta_l}{L}\hat \mu\;,\;\;%
		\rho = \hat \rho (\rho_l + \rho_g)\;,\;\;
		\eta = \hat \eta (\eta_l + \eta_l)\;,\;\;
g = \frac{U^2}{L}\hat g\;,
\end{align}
and apply the following nondimensional groups:
\begin{align}
\Rey=\frac{\rho_l U L}{\eta_l}\;,\;\;
\Ca=\frac{\eta_l U}{\sigma}\;,\;\;
\Bo=\frac{\rho_l g L^2}{\sigma}\;,\;\;
\A_\rho=\frac{\rho_l - \rho_g}{\rho_l + \rho_g}\;,\;\;
\A_\eta=\frac{\eta_l - \eta_g}{\eta_l + \eta_g}\;,\;\;
\Cn=\frac{\epsilon}{L}\;,\;\;
\Pe_\epsilon=\frac{\epsilon L U}{b\sigma}\;,
\end{align}
where $L=\delta_{nu}$ and $U=\bar v_{nu}$ are the Nusselt film height and the mean film velocity, respectively.
Both are calculated for a specific Reynolds number $\Rey$ and the inclination angle $\alpha$ measured from the plate to the horizontal with:
\begin{align}
		\delta_{nu}=\sqrt[3]{\frac{3 \eta_l^2 \Rey}{\rho_l^2 g \sin\alpha}}\;, \label{eqn:delta_nu} \\
		\bar v_{nu}= \frac{\delta_{nu}^2 \rho_l g \sin\alpha}{3 \eta_l}\;. \label{eqn:v_nu}
\end{align}

In this way the physical system is characterized by the 
\begin{itemize}
	\item Reynolds number $\Rey$ (inertial over viscous forces in the film), 
	\item Capillary number $\Ca$ (viscous drag in the film over surface tension forces between film and gas), 
	\item Bond number $\Bo$ (gravitational force in the film over surface tension forces between film and gas),
	\item and Atwood numbers $\A_\rho$ and $\A_\eta$ (density and viscosity ratios between film and gas).
\end{itemize}
The Cahn number $\Ca$ and the Peclet number $\Pe_\epsilon$ stem from the Cahn--Hilliard approach and describe the dynamics of the diffuse interface.

We obtain the following nondimensionalized CHNS system\footnote{For better readability we omit the $\hat\_$ above all scaled variables and operators from now on. If not otherwise noted, starting from~\cref{eq:M:1_NS1} all variables are scaled and dimensionless.}: 
\begin{align}
		\rho\partial_t v + ((\rho v + J)\cdot\nabla) v -\Rey^{-1}\mbox{div}\left(2\eta Dv\right) + \nabla p &= -\Rey^{-1}\varphi\nabla \mu + \Bo \Ca^{-1} \Rey^{-1} \rho g && \mbox{ in } \Omega, \label{eq:M:1_NS1}\\
-\mbox{div}(v) &= 0 && \mbox{ in } \Omega,\label{eq:M:2_NS2}\\
\partial_t \varphi + v \cdot\nabla \varphi - \Ca\Cn\Pe_\epsilon^{-1}\Delta \mu &= 0 && \mbox{ in } \Omega,\label{eq:M:3_CH1}\\
-\Cn^2\Delta \varphi + W^\prime(\varphi) &= \Ca\Cn\mu && \mbox{ in } \Omega,\label{eq:M:4_CH2}
\end{align}
with $J := -\Ca\Cn\Pe_\epsilon^{-1}\frac{\partial\rho}{\partial\varphi}\nabla \mu$ and
\begin{align}
	\rho = \frac{1}{2}\left(1 + \varphi \A_\rho)\right)\;,\;\;\;
	\eta = \frac{1}{2}\left(1 + \varphi \A_\eta)\right)\;.
\end{align}

\section{Numerical Method}\label{sec:numerics}

\subsection{Discretization}\label{sec:disc}
For a practical implementation in a finite element scheme a time grid $0 = t_0 < t_1 < \ldots<t_{m-1} < t_m< \ldots <t_M = T$ on $I = [0,T]$ with non-equidistant step size $\tau^m>0$ is introduced.
Further, a triangulation $\mathcal T_h$ of the domain into cells $T_i$ is introduced such that $\mathcal T_h = \bigcup_{i=1}^{N}T_i$ covers the domain. 
The specific meshing is discussed in~\cref{sec:mesh}.
On $\mathcal T_h$ we introduced piecewise linear Lagrange finite elements $V_1 = \mathcal{P}_1$ for $\varphi_h$, $\mu_h$ and $p_h$ and the triangular/tetrahedral Mini element $V_M = \mathcal{P}_1\bigoplus \mathcal{B}_{1+d}$, denoting the space of linear polynomials enriched by a cubic bubble function, for $v_h$.
For the derivation of the weak form as well as the proof of energy stability and thermodynamic consistency we refer to~\cite{Bonart2019b}.

Given 
$\varphi^{m-1} \in V_1$, 
$\mu^{m-1} \in V_1$, and
$v^{m-1} \in V_M$, 
find 
$\varphi^m_h \in V_1$, 
$\mu^m_h \in V_1$,
$p^m_h \in V_1$ and
$v^m_h \in V_M$,
such that for all 
$w \in V_M$,
$q \in V_1$,
$\Phi \in V_1$, and
$\Psi \in V_1$
the following equations hold:
\begin{align}
  \frac{1}{\tau^m}\left(\frac{\rho^m+\rho^{m-1}}{2} v^m_h -\rho^{m-1}v^{m-1},w\right)\nonumber\\
  + a(\rho^{m-1}v^{m-1} + J^{m-1},v^m_h,w)
		+ (\Rey^{-1} 2\eta^{m-1}Dv^m_h,Dw) - (\mbox{div} w,p^m_h)\nonumber\\
		+(\Rey^{-1} \varphi^{m-1}\nabla \mu^m_h,w) -\Bo \Ca^{-1} \Rey^{-1}(g\rho^{m-1},w) &= 0,
    \label{eq:S:1_NS_1}\\
    -(\mbox{div} v^m_h,q) &= 0, \label{eq:S:2_NS_2}\\
   \frac{1}{\tau^m}(\varphi_h^{m} - \varphi^{m-1},\Psi)
	-\left ( \varphi^{m-1}v^{m-1}, \nabla \Psi \right )
		+ \left (\Rey^{-1} \frac{\tau^m|\varphi^{m-1}|^2 }{\rho^{m-1}} \nabla \mu^m_h, \nabla \Psi\right )\nonumber\\
		+(\Ca\Cn\Pe_\epsilon^{-1}\nabla \mu^m_h,\nabla \Psi) &= 0 \label{eq:S:3_CH1},\\
  (\Cn^2\nabla \varphi^m_h,\nabla \Phi)
   +(W^\prime(\varphi^{m-1}) + S_W(\varphi^{m}_h - \varphi^{m-1}),\Phi)  
   - (\Ca\Cn\mu^m_h,\Phi) &= 0,
    \label{eq:S:4_CH2}
\end{align}
with 
$J^{m-1} := -\Pe_\epsilon^{-1}\frac{\partial \rho}{\partial\varphi}(\varphi^{m-1})\nabla \mu^{m-1}$,
$\rho^{m-1} := \rho(\varphi^{m-1})$, and $\eta^{m-1} := \eta(\varphi^{m-1})$.
We decoupled the Navier--Stokes equation and the Cahn--Hilliard equation by using an augmented velocity field in \cref{eq:S:3_CH1}, see \cite{Minjeaud2013,Gruen2016}.
In this way, one can first solve \cref{eq:S:3_CH1,eq:S:4_CH2} and thereafter \cref{eq:S:1_NS_1,eq:S:2_NS_2}. 
Furthermore, for $W^\prime$ a stabilized linear scheme was applied, where $S_W$ is a suitable stabilization parameter.
Note that this decoupled and linearized scheme is also energy stable and thermodynamically consistent, see~\cite{Bonart2019b,2015-ShenYangYu-EnergyStableSchemesForCHMCL-Stabilization}.

To control the time step size we used a simple and straightforward strategy.
After every time step iteration we calculated the minimal time step based on two distinct Courant numbers $\Co_v$ and $\Co_\varphi$.
The minimal time step $\tau_v^m$ was calculated following the well-known Courant relation, see~\cite{Ferziger2008}.
Furthermore, to include the movement of the interface into the time step consideration, we follow~\cite{Aland2014} and replaced the velocity $v^{m-1}$ with the phase field velocity 
\begin{align}
		\frac{\partial_t \varphi}{|\nabla \varphi|}\approx \frac{\varphi^{m-1} - \varphi^{m-2}}{\tau^{m-1}|\nabla \varphi^{m-1}|}
\end{align}
to obtain $\tau_\varphi^m$.
The next time step $\tau^m$ was chosen as the minimum of $\tau_v^m$ and $\tau_\varphi^m$ as well as $\tau_{max}$ to restrict the time step size to reasonable values especially at the beginning of the simulations:
\begin{align}
		\tau^m 
			   = \min(\min(\tau_v^m), \min(\tau_\varphi^m), \tau_{max})
			   = \min\left(\min\left(\frac{\Co_v h}{|v^{m-1}|}\right), \max\left(\frac{\Co_\varphi \tau^{m-2}h|\nabla \varphi^{m-1}|}{|\varphi^{m-1}-\varphi^{m-2}|}\right), \tau_{max} \right) \;.
\end{align}

\subsection{Solver}
We implemented the solution scheme given in~\cref{sec:disc} in Python3 using the finite element library FEniCS 2019.1.0~\cite{AlnaesBlechta2015a,fenics_book}.
For the solution of the arising linear systems and subsystems the software suite PETSc 3.8.4~\cite{petsc_webpage, petsc-user-ref, petsc-efficient} was applied.
At each time step we first solved the CH system and thereafter the NS systems.
The CH system was solved using the direct linear solver MUMPS 5.1.1 \cite{mumps_1, mumps_2}.
Note that the naive usage of a Krylov method for unsymmetric systems, e.g., GMRES, preconditioned by a simple Gauss-Seidel or successive over relaxation method to solve the CH system results in a lot of outer iterations. 
We refer to~\cite{Boyanova2012,Bosch2018} for efficient preconditioners for the CH system.

For the NS system we had to solve at every time step a linear system with the linear operator $G$:
\begin{align}
	G_{NS} &= \begin{pmatrix} 
			\frac{1}{\tau}\left(\frac{\rho^m+\rho^{m-1}}{2} v^m_h,w\right) + a(\rho^{m-1}v^{m-1} + J^{m-1},v^m_h,w) + (\Rey^{-1} 2\eta^{m-1}Dv^m_h,Dw)& - (\mbox{div} w,p^m_h) \\
			-(\mbox{div} v^m_h,q) & 0
		\end{pmatrix} \\
		&= \begin{pmatrix}
				A &B^T\\
				B &0
			\end{pmatrix}
\end{align}
Solving this large-scale saddle point system is computationally very expensive.
Furthermore, large meshes forbid the usage of direct linear solvers.
Therefore, we applied PETSc's GMRES method preconditioned from the right.
As preconditioner an upper triangular block preconditioner for Oseen type problems was used:
\begin{align}
		P_{NS} = \begin{pmatrix}
				\hat A &B^T\\
				0 & -\hat S
		\end{pmatrix}\;,
\end{align}
where $S$ is an approximation of the Schur complement given by the pressure-convection-diffusion (PCD) preconditioner 
\begin{align}
		S^{-1} =R_p^{-1} + M_p^{-1} (I + K_p A_p^{-1}). \label{eqn:pcd}
\end{align}
For details about this preconditioner we refer to~\cite{Blechta2019,Olshanskii2007,Elman2008}.
Recently, this preconditioner was generalized to two-phase flows~\cite{Bootland2019}.
In this work, we used the following expressions for the matrices occurring in~\cref{eqn:pcd}:
\begin{align}
	M_p &= \frac{\Rey}{\eta^{m-1}}(p_h^m,q)\;, \\
	K_p &= \frac{\Rey}{\eta^{m-1}}a(\rho^{m-1}v^{m-1} + J^{m-1},p^m_h,q)\;,\\
	A_p &= (\nabla p, \nabla q)\;, \\
	R_p &= B \left(\text{diag}(M_v)\right)^{-1} B^T\;,\\
	M_v &= \frac{1}{\tau}\left(\frac{\rho^m+\rho^{m-1}}{2} v^m_h, w\right )
\end{align}
where $M_p$, $M_v$ are scaled pressure and velocity mass matrices, respectively, $K_p$ is a scaled pressure convection matrix, and $A_p$ is the pressure Laplacian. 

For details on the implementation of the preconditioner we refer to the documentation of FENaPACK~\url{https://fenapack.readthedocs.io}.
Besides the default options in FENaPACK and PETSc the Richardson method was applied together with algebraic multigrid provided by Hypre as preconditioner for the inversion of $A$, $R_p$ and $A_p$. 
For the inversion of $M_p$ the preconditioned Chebyshev iterative method together with a Jacobi preconditioner was used.

The model and the solution scheme as well as our implementation has been extensively validated.
We obtained accurate results against the well-known rising bubble benchmark by~\citet{Hysing2009}, see~\cite{Bonart2019b}.
Flows involving moving contact lines were validated against analytical solutions of spreading and pinned droplets in~\cite{Bonart2019a}.
Again we accomplished a very good agreement.
We obtained an almost perfect accordance with the analytical Nusselt film solution in~\cite{Bonart2018a}.
For a validation involving film flows over corrugations, we matched our results with the experiments reported by~\cite{Wierschem2010} and the simulations by~\cite{Dietze2019}, see~\ref{sec:validation}.

\section{Simulation Case and Mesh}\label{sec:mesh}
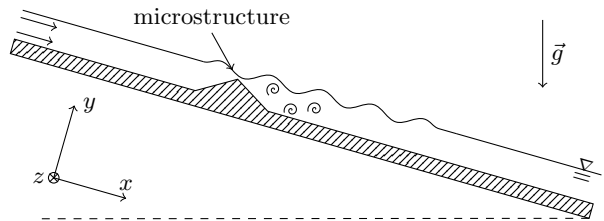
\begin{figure}[h!]
\centering
\begin{minipage}[]{0.60\textwidth}
\begin{tikzpicture}
	\draw [pattern=north east lines,rotate around={-15.9:(0,0)}] (2,0) -- (2,0.2) -- (4.5,0.2) -- (5,0.5) -- (5.5,0.2) -- (10,0.2) -- (10,0) --cycle;
	\draw[<-,rotate around={-15.9:(0,0)}] (4.9,0.6) -- (4.5,1) node[above]{microstructure};
	\draw [thin,rotate around={-15.9:(0,0)}] (2,0.6) -- (4.5,0.6);
	\draw [decorate, decoration={snake, amplitude=2pt, segment length=20pt},rotate around={-15.9:(0,0)}] (4.5,0.6) -- (8.0,0.6);
	\draw [thin,rotate around={-15.9:(0,0)}] (8,0.6) -- (10,0.6) node[left,rotate around={-15.9:(0,0)}] {$\overset{\nabla}{=}$} ;
	\draw [->,rotate around={-15.9:(0,0)}](2,0.3) -- ++(0.5,0);
	\draw [->,rotate around={-15.9:(0,0)}](2,0.5) -- ++(0.5,0);
	\draw [thin,dashed,rotate around={-15.9:(0,0)}] (10,0) -- (3,-2);
	\draw[->,rotate around={-15.9:(2.5,-2.2)}] (2.5,-2.2) -- ++(1,0) node[above]{$x$};
	\draw[->,rotate around={-15.9:(2.5,-2.2)}] (2.5,-2.2) -- ++(0,1) node[right]{$y$};
	\draw (2.5,-2.2) circle (2pt) node[left]{$z$};
	\draw (2.5,-2.2) node[cross=2pt] {};
	\spiral[rotate around={-15.9:(0,0)}](5.5,0.5)(-115:2:.1);
	\spiral[rotate around={-15.9:(0,0)}](5.8,0.3)(-145:2:.1);
	\spiral[rotate around={-15.9:(0,0)}](6.1,0.4)(-145:2:.1);
	\draw[->] (9,-0.1) --  node[right]{$\vec g$}(9,-1);
\end{tikzpicture}
\end{minipage}
\caption{Illustration of the simulation domain. Exemplary, a triangular microstructure with sharp corner is displayed.}
\label{fig:testcase}
\end{figure}

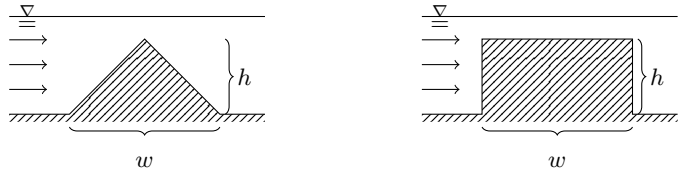
\begin{figure}[h!]
\centering
\begin{minipage}[]{0.3\textwidth}
\begin{tikzpicture}
	\draw (0,0) -- ++(0.8,0) -- ++(1,1) -- ++(1,-1) -- ++(0.6,0);
	\fill[thin,pattern=north east lines] (0,0) -- ++(0.8,0) -- ++(1,1) -- ++(1,-1) -- ++(0.6,0) -- ++(0,-0.1) -- ++(-3.4,0); 
	\draw (0,1.3) node[right] {$\overset{\nabla}{=}$} -- ++(3.4,0);
	\draw[arrows=->](0,0.99) -- ++(0.5,0);
	\draw[arrows=->](0,0.66) -- ++(0.5,0);
	\draw[arrows=->](0,0.33) -- ++(0.5,0);
  	\draw [decorate,decoration={brace,amplitude=3pt,raise=4pt,mirror},xshift=-2pt] (2.8,0) -- ++(0,1) node [midway,xshift=+0.4cm] {$h$};
	\draw [decorate,decoration={brace,amplitude=3pt,raise=4pt,mirror},yshift=2pt] (0.8,-0.1) -- ++(2,0) node [midway,yshift=-0.6cm] {$w$};
\end{tikzpicture}
\end{minipage}
\begin{minipage}[]{0.3\textwidth}
\begin{tikzpicture}
  	\draw (0,0) -- ++(0.8,0) -- ++(0,1) -- ++(2,0) -- ++(0,-1) -- ++(0.6,0);
	\fill[thin,pattern=north east lines] (0,0) -- ++(0.8,0) -- ++(0,1) -- ++(2,0) -- ++(0,-1) -- ++(0.6,0) -- ++(0,-0.1) -- ++(-3.4,0); 
	\draw (0,1.3) node[right] {$\overset{\nabla}{=}$} -- ++(3.4,0);
	\draw[arrows=->](0,0.99) -- ++(0.5,0);
	\draw[arrows=->](0,0.66) -- ++(0.5,0);
	\draw[arrows=->](0,0.33) -- ++(0.5,0);
  	\draw [decorate,decoration={brace,amplitude=3pt,raise=4pt,mirror},xshift=-2pt] (2.8,0) -- ++(0,1) node [midway,xshift=+0.4cm] {$h$};
	\draw [decorate,decoration={brace,amplitude=3pt,raise=4pt,mirror},yshift=2pt] (0.8,-0.1) -- ++(2,0) node [midway,yshift=-0.6cm] {$w$};
\end{tikzpicture}
\end{minipage}
\caption{Illustration of the microstructures.}
\label{fig:microstructures}
\end{figure}

In~\cref{fig:testcase} the simulation domain is illustrated. 
Exemplary, a triangular obstacle with base length and height of $h=0.75$ is shown. 
Due to the nondimensionalization, see~\cref{sec:nondim}, $h$ is always given relative to the film height.
The geometries of the two examined microstructures are depicted in~\cref{fig:microstructures}.
We chose triangular and rectangular as representative structures as they are the most simple forms with sharp corners.
The solid surface was tilted to the horizontal with angle $\alpha$.
The flow entered the simulation domain from the left and emitted at the right.
Periodic boundary conditions were applied on the left and right side of the domain. 
The flow was purely driven by gravity with the gravitational acceleration constant $g$. 
On the bottom wall no-slip boundary conditions were used. 
The size of the symmetric domain was $8L\times4L$.

\begin{figure}[h!]
	\centering
		\includegraphics[width=0.5\linewidth]{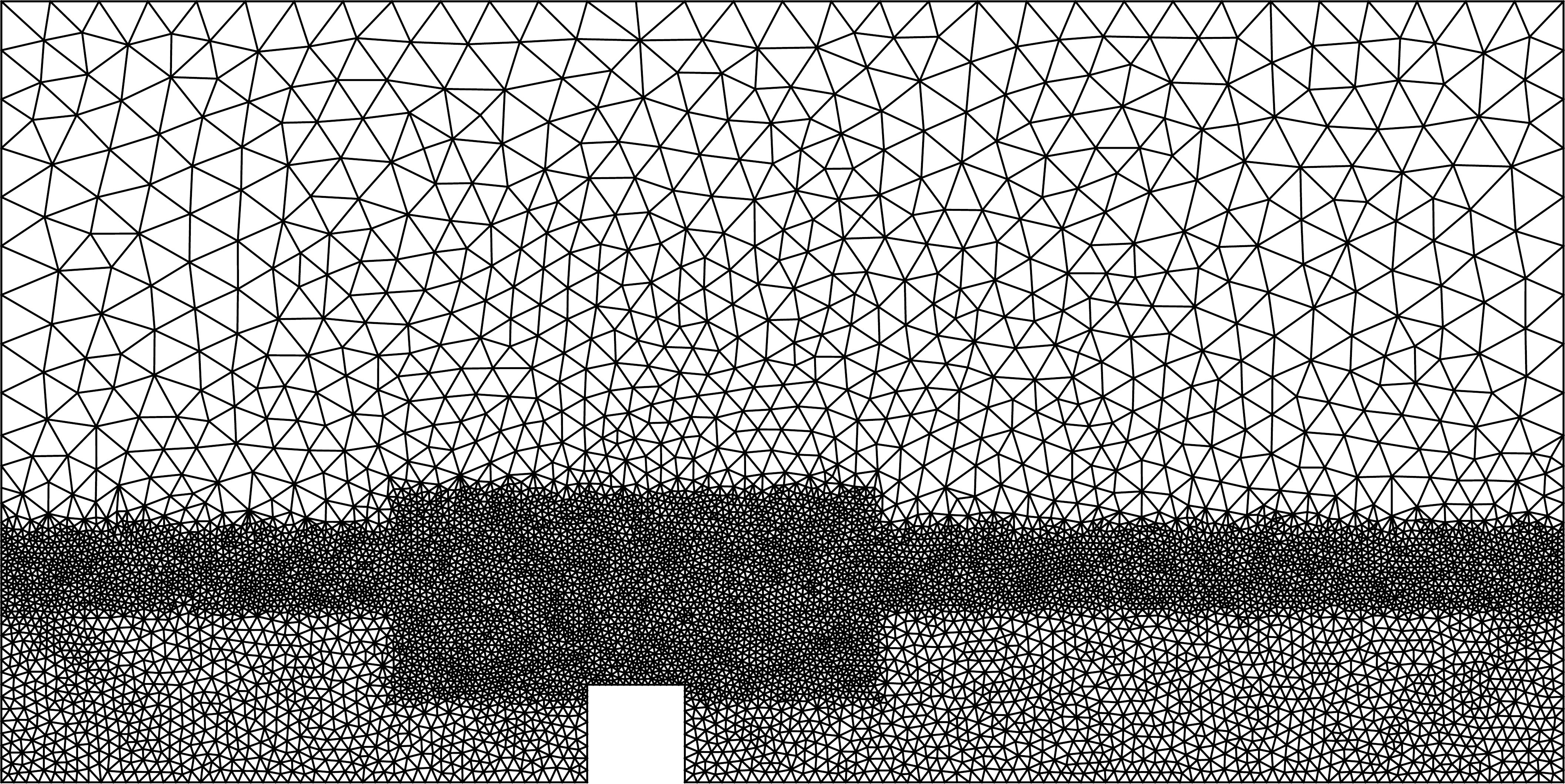}
		\caption{Two-dimensional, unstructured, periodic, triangular mesh used in the simulations of the rectangular microstructures. It has three refinement zones: interface, film and obstacle. The size of the bounding box is $8L\times4L$. This mesh consists of around 1594 vertices and 3075 elements.}
	\label{fig:mesh}
\end{figure}

\Cref{fig:mesh} displays the mesh used in the simulations.
The two-dimensional, unstructured, triangular, periodic meshes were generated using Gmsh~\cite{Geuzaine2009}.
The background cell diameter was of size $h_{outer} = 32\Cn/5$.
The area around the interface was resolved with $h_{interface} = 4\Cn/5$, which resulted in around five cells over the interfacial thickness.
If was found by~\cite{Cai2015} that this is sufficient to accurately capture the dynamics of the interface.
The film was resolved with $h_{film} = 2 h_{interface}$.
In this way, the mesh consisted of 1594 vertices and 3075 elements.
Using the scheme from~\cref{sec:disc} and $\Cn=0.04$ the meshing resulted in around 16,000 and 55,000 degrees of freedom for the CH and the NS system,  respectively.

\section{Film Flow Over Microstructures}\label{sec:results}
\subsection{Setup}
To compare the stability of the liquid films, we performed simulations with Reynolds numbers $\Rey$ between 10 and 70.
Following~\cite{Dietze2019} we assume, that two-dimensional structures can be represented with infinite depth normal to the main flow direction in a two-dimensional simulation.
We initialized the simulation for a specific value of $\Rey$ with a smooth film of height $\delta_{Nu}$ calculated from~\cref{eqn:delta_nu}.
The nondimensional parameters are listed in~\cref{tab:parameters}.
The initial velocity of the film and the overflowing gas phase was zero, i.e., the liquid film as well as the gas phase were at rest.
This corresponds to an initial inclination angle $\alpha=\SI{0}{\degree}$.
At the very beginning of the simulation the plate was flipped to an inclination of $\alpha=\SI{8}{\degree}$.
Due to the periodic domain the distance between the simulated microstructure and the subsequent microstructure was $>7\delta$.
The simulations were performed until a final time of $T=\SI{2}{\second}$ or until a steady-state was reached.
\begin{table}[!h]
		\centering
\begin{tabular}{ccccccc}
\toprule
		$\Rey$&$\Ca$&$\Bo$&$\A$&$\A_\eta$&$\Cn$&$\Pe_\epsilon$\\
\midrule
		10& 0.04& 0.0026& \multirow{7}{*}{0.99}&\multirow{7}{*}{0.99} & \multirow{7}{*}{0.04} & 107 \\
		20& 0.07& 0.0041& &  & & 169 \\
		30& 0.09& 0.0054& &  & & 222 \\
		40& 0.11& 0.0066& &  & & 269 \\
		50& 0.13& 0.0076& &  & & 313 \\
		60& 0.15& 0.0086& &  & & 353 \\
		70& 0.16& 0.0095& &  & & 391 \\
\midrule
\bottomrule
\end{tabular}
\caption{Nondimensional parameters used in the film flow simulations.}
\label{tab:parameters}
\end{table}

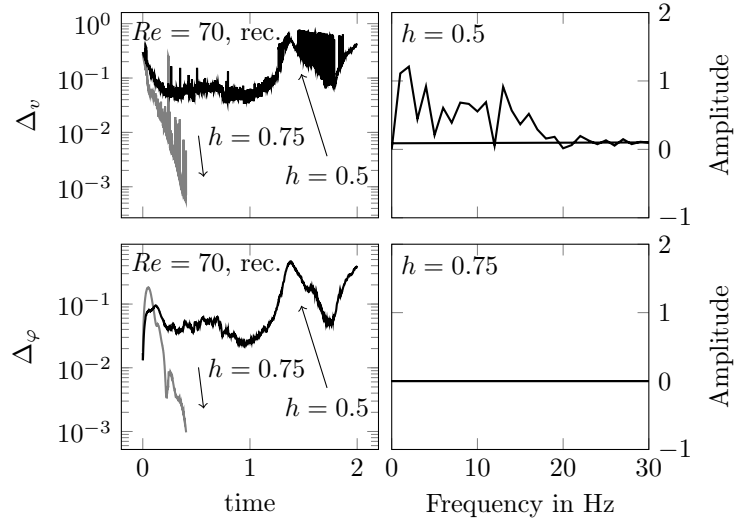
\begin{figure}[]
\tikzsetnextfilename{steadystate}
\centering
\begin{tikzpicture}
	\centering
	\begin{groupplot}[
			group style = {
				group size = 2 by 2
				,vertical sep=10pt,horizontal sep=5pt
				,xlabels at=edge bottom
               	,xticklabels at=edge bottom
			},
    		width = 5cm
			,xlabel near ticks
			,ylabel near ticks
			]
			\nextgroupplot[ymode=log, ylabel=$\Delta_v$]
				\node[right] at (rel axis cs:0,0.9){$\Rey=70$, rec.};
				\addplot[thick, color=gray] table [x expr=\coordindex/10000*10, y=Dv] {data/2d/log/steadystate.dat};
				\draw[->] (rel axis cs:0.3,0.4)node[right]{$h=0.75$} to (rel axis cs:0.32,0.2);

				\addplot[thick] table [x expr=\coordindex/10000*10, y=Dv] {data/2d/log/steadystate2.dat};
				\draw[->] (rel axis cs:0.8,0.3)node[below]{$h=0.5$} to (rel axis cs:0.7,0.7);
			\nextgroupplot[ymin=-1, ymax=2,xmin=0,xmax=30,ylabel near ticks, yticklabel pos=right,ylabel={Amplitude}]
				\node[right] at (rel axis cs:0,0.9){$h=0.5$};
				\addplot[thick, black] table [x index = 1, y index = 0] {data/2d/waves/2d_rec_h0.5_Re70.0_Cn0.04_fft.dat};
			\nextgroupplot[ymode=log, ylabel=$\Delta_\varphi$,xlabel=time]%
				\node[right] at (rel axis cs:0,0.9){$\Rey=70$, rec.};
				\addplot[thick, color=gray] table [x expr=\coordindex/10000*10, y=Dphi] {data/2d/log/steadystate.dat};
				\draw[->] (rel axis cs:0.3,0.4)node[right]{$h=0.75$} to (rel axis cs:0.32,0.2);

				\addplot[thick] table [x expr=\coordindex/10000*10, y=Dphi] {data/2d/log/steadystate2.dat};
				\draw[->] (rel axis cs:0.8,0.3)node[below]{$h=0.5$} to (rel axis cs:0.7,0.7);
			\nextgroupplot[ymin=-1, ymax=2,xlabel={Frequency in \si{\hertz}},ylabel={Amplitude},xmin=0,xmax=30,ylabel near ticks, yticklabel pos=right]
				\node[right] at (rel axis cs:0,0.9){$h=0.75$};
				\addplot[thick, black] table [x index = 1, y index = 0] {data/2d/waves/2d_rec_h0.75_Re70.0_Cn0.04_fft.dat};
		\end{groupplot}
	\end{tikzpicture}
		\caption{Development of the steady state criteria and the corresponding amplitude spectra at $x=3$ for 1 to \SI{2}{\second}.}	
\label{fig:steadystate}
\end{figure}

To decide if a steady-state was reached, we looked at the change of the key variables velocity $v$ and phase field $\varphi$ over one time step~\cite{Aland2014}:
\begin{align}
		\Delta_v = \int_\Omega \frac{|v_x^m - v_x^{m-1}|}{\tau^m}dx\;,\;\;\; \Delta_\varphi = \int_\Omega \frac{|\varphi^m - \varphi^{m-1}|}{\tau^m}dx\;.
\end{align}
Exemplarily, in~\cref{fig:steadystate} the steady state criteria for two rectangular structures with $h=0.75$ and $h=0.5$ are plotted over time (left).
Furthermore, we show the corresponding amplitude spectra measured at $x=3$ for 1 to \SI{2}{\second}.
It is clearly evident, that the large rectangular microstructure with $h=0.75$ led to very small changes from one time step to another in the first \SI{0.5}{\second} (gray lines).
The steady-state was reached very quickly, no waves were formed and the corresponding amplitude spectrum (bottom right panel in~\cref{fig:steadystate}) is zero.
In contrast, the waves for $h=0.5$ and $\Rey=70$ completely prevented a steady-state (black lines in~\cref{fig:steadystate}).

\subsection{Results}
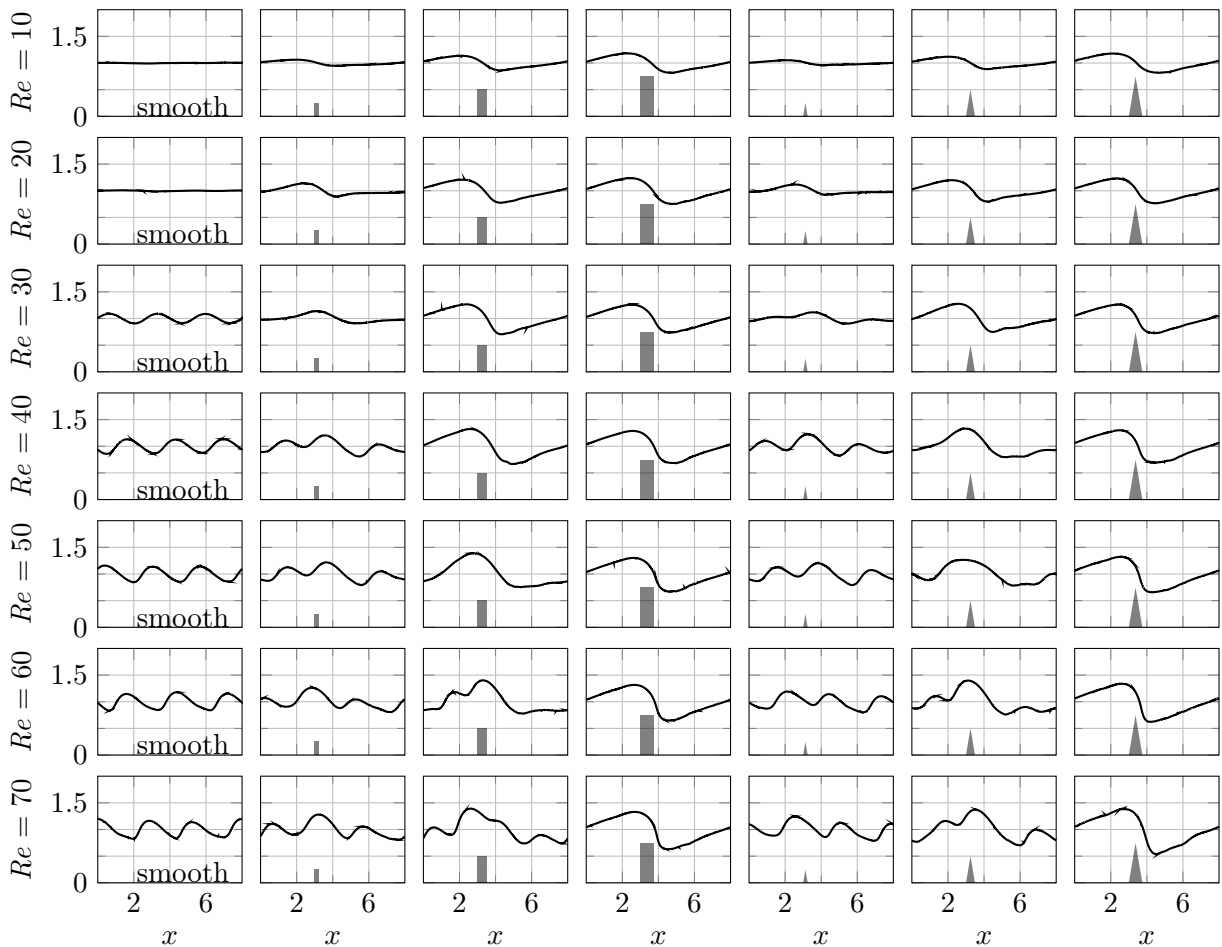
\begin{figure}
\tikzsetnextfilename{film2d}
\centering
\begin{tikzpicture}
	\centering
	\begin{groupplot}[
			group style = {
				group size = 7 by 7
				,vertical sep=8pt,horizontal sep=7pt
				,ylabels at=edge left
				,xlabels at=edge bottom
               	,yticklabels at=edge left 
				,xticklabels at=edge bottom
			},
    		width = 3.5cm, height = 3cm
			,xlabel near ticks
			,ylabel near ticks
			, xmin=0, xmax=8
			, ymin=0, ymax=2.0
			, ytick={0,1.5}
		    , xtick={2,6}
			, minor y tick num=2,
			, minor x tick num=1,
			, xlabel={$x$}
			, grid=both
			]
			\pgfplotsinvokeforeach {10,20,30,40,50,60,70}
				{ 
					\nextgroupplot[ylabel={$\Rey=#1$}] 
						\node[right] at (rel axis cs:0.2,0.1){smooth};
						\addplot[thick, smooth] plot file{data/2d/isolines/2d_smooth_h0.0_Re#1.0_Cn0.04_phi0.dat};
					\nextgroupplot[] 
						\addplot[thick, smooth] plot file{data/2d/isolines/2d_rec_h0.25_Re#1.0_Cn0.04_phi0.dat};
						\fill[opacity=0.5, black] (3, 0) -- (3,0.25) -- (3.25, 0.25) -- (3.25, 0) -- cycle;
					\nextgroupplot[] 
						\addplot[thick, smooth] plot file{data/2d/isolines/2d_rec_h0.5_Re#1.0_Cn0.04_phi0.dat};
						\fill[opacity=0.5, black] (3, 0) -- (3,0.5) -- (3.5, 0.5) -- (3.5, 0) -- cycle;
					\nextgroupplot[] 
						\addplot[thick, smooth] plot file{data/2d/isolines/2d_rec_h0.75_Re#1.0_Cn0.04_phi0.dat};
						\fill[opacity=0.5, black] (3, 0) -- (3,0.75) -- (3.75, 0.75) -- (3.75, 0) -- cycle;
					\nextgroupplot[] 
						\addplot[thick, smooth] plot file{data/2d/isolines/2d_tri_h0.25_Re#1.0_Cn0.04_phi0.dat};
						\fill[opacity=0.5, black] (3, 0) -- (3.125,0.25) -- (3.25, 0) -- cycle;
					\nextgroupplot[] 
						\addplot[thick, smooth] plot file{data/2d/isolines/2d_tri_h0.5_Re#1.0_Cn0.04_phi0.dat};
						\fill[opacity=0.5, black] (3, 0) -- (3.25,0.5) -- (3.5, 0) -- cycle;
					\nextgroupplot[] 
						\addplot[thick, smooth] plot file{data/2d/isolines/2d_tri_h0.75_Re#1.0_Cn0.04_phi0.dat};
						\fill[opacity=0.5, black] (3, 0) -- (3.375,0.75) -- (3.75, 0) -- cycle;
				}
		\end{groupplot}
\end{tikzpicture}
		\caption{Shape of the liquid film interface at $\tau=\SI{2}{\second}$ for different values of $\Rey$. The domain is compressed and cut off at a height of 2.0. All dimensions are normalized with the respective film thickness $\delta$.}	
\label{fig:res:film2d}
\end{figure}

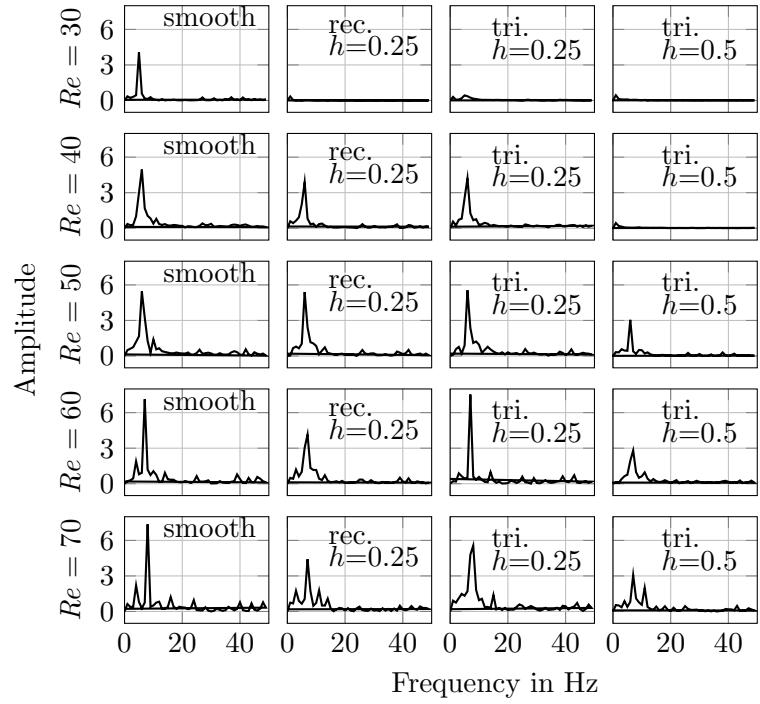
\begin{figure}[]
\tikzsetnextfilename{film2d_fft}
\centering
\begin{tikzpicture}
	\centering
	\begin{groupplot}[
			group style = {
				group size = 4 by 5
				,vertical sep=8pt,horizontal sep=7pt
				,ylabels at=edge left
				,xlabels at=edge bottom
               	,yticklabels at=edge left 
				,xticklabels at=edge bottom
			},
    		width = 3.5cm, height = 3cm
			,xlabel near ticks
			,ylabel near ticks
			, xmin=0, xmax=50
			, ymin=-1, ymax=8
			, ytick={0,3,6}
			, grid=both
			]
			\pgfplotsinvokeforeach {30,40,50,60,70}
				{ 
					\nextgroupplot[ylabel={$\Rey=#1$}] 
						\node[right] at (rel axis cs:0.2,0.9){smooth};
						\addplot[thick, black] table [x index = 1, y index = 0] {data/2d/waves/2d_smooth_h0.0_Re#1.0_Cn0.04_fft.dat};
					\nextgroupplot[] 
						\node[align=left, execute at begin node=\setlength{\baselineskip}{1ex}] at (rel axis cs:0.6,0.7){rec.\\$h{=}0.25$};
						\addplot[thick, black] table [x index = 1, y index = 0] {data/2d/waves/2d_rec_h0.25_Re#1.0_Cn0.04_fft.dat};
					\nextgroupplot[] 
						\node[align=left, execute at begin node=\setlength{\baselineskip}{1ex}] at (rel axis cs:0.6,0.7){tri.\\$h{=}0.25$};
						\addplot[thick, black] table [x index = 1, y index = 0] {data/2d/waves/2d_tri_h0.25_Re#1.0_Cn0.04_fft.dat};
					\nextgroupplot[] 
						\node[align=left, execute at begin node=\setlength{\baselineskip}{1ex}] at (rel axis cs:0.6,0.7){tri.\\$h{=}0.5$};
						\addplot[thick, black] table [x index = 1, y index = 0] {data/2d/waves/2d_tri_h0.5_Re#1.0_Cn0.04_fft.dat};
				}
		\end{groupplot}
    		\node[above, rotate=90] at (current bounding box.west) {Amplitude};
			\node[below right] at (current bounding box.south) {Frequency in \si{\hertz}};
\end{tikzpicture}
		\caption{Amplitude spectra for some liquid films calculated using a discrete Fourier transformation. Stable films with zero amplitude spectra are omitted. The data for the DFT was extracted from the isolines, see~\cref{fig:res:film2d}, at $x=3$ (right before the microstructures) for 1 to \SI{2}{\second}}
\label{fig:res:film2d_fft}
\end{figure}

We observed a completely smooth film for low $\Rey$ and an onset of waves for intermediate $\Rey$.
For larger $\Rey$ we even saw a highly distorted, almost chaotic, film with irregular waves, see below or~\cref{fig:res:film2d} and~\cref{fig:res:film2d_fft}.
Qualitatively, we decided that a film is unstable if any waves were formed at all after at most \SI{2}{\second}.
A second criteria is that the amplitude spectrum is low without any pronounced peaks.
In this way, we do not take a single, stagnant bump due to retaining as a sign for an unstable film (for example, we consider the films in the fourth column from~\cref{fig:res:film2d} as stable).

In~\cref{fig:res:film2d} the film surface obtained for two microstructures (rectangular and triangular) with height and width $h=w=0.75\delta$, $h=w=0.5\delta$ and $h=w=0.25\delta$ are displayed for Reynolds numbers $\Rey$ ranging from 10 to 70.
For comparison the film over a smooth surface is shown in the first column.
The film surfaces (depicted as solid black lines) were extracted from the simulation data as the isolines where $\varphi=0$.
The microstructures are illustrated as gray insets in~\cref{fig:res:film2d}. 
To gain more insight into the waves~\cref{fig:res:film2d_fft} shows amplitude spectra for a choice of configurations.
The amplitudes were calculated using a discrete Fourier transformation (DFT).
The data for the DFT was extracted from the isolines, see~\cref{fig:res:film2d}, at $x=3$ (right before the microstructures) for 1 to \SI{2}{\second}.

The film over the smooth surface did not show any deformation for low numbers of $\Rey$.
As expected, waves started to form for $\Rey\ge30$, which got more and more pronounced with increasing $\Rey$.
However, even for $\Rey=70$, the waves stayed relatively regular.
The amplitude spectra showed a pronounced, sharp peak for all $\Rey$, which is a strong sign for regular waves.

\begin{figure}[]
\tikzsetnextfilename{film2d_time}
\centering
\begin{tikzpicture}
	\centering
	\begin{groupplot}[
			group style = {
				group size = 2 by 6
				,vertical sep=8pt,horizontal sep=7pt
				,ylabels at=edge left
				,xlabels at=edge bottom
               	,yticklabels at=edge left 
				,xticklabels at=edge bottom
			},
    		width = 3.5cm, height = 3cm
			,xlabel near ticks
			,ylabel near ticks
			, xmin=0, xmax=8
			, ymin=0, ymax=2.0
			, ytick={0,1.5}
            , xtick={2, 6}
			, minor y tick num=2,
		    , minor x tick num=1,
			, xlabel={$x$}
			, grid=both
			]
				\pgfplotsinvokeforeach {1.5,1.6,1.7,1.8,1.9,2.0}
				{
					\nextgroupplot[ylabel={$t=\SI{#1}{\second}$}] 
						\addplot[thick, smooth] plot file{data/2d/rec_time/2d_rec_h0.25_Re70.0_Cn0.04_phi0_t#1.dat};
						\fill[opacity=0.5, black] (3, 0) -- (3,0.25) -- (3.25, 0.25) -- (3.25, 0) -- cycle;
					\nextgroupplot[] 
						\addplot[thick, smooth] plot file{data/2d/rec_time/2d_rec_h0.5_Re70.0_Cn0.04_phi0_t#1.dat};
						\fill[opacity=0.5, black] (3, 0) -- (3,0.5) -- (3.5, 0.5) -- (3.5, 0) -- cycle;
				}
		\end{groupplot}
\end{tikzpicture}
		\caption{Shape of the liquid film interface over rectangular microstructures at different instances of time for $\Rey=70$. The domain is compressed and cut off at a height of 2.0. All dimensions are normalized with the respective film thickness $\delta$.}	
\label{fig:res:film2d_time}
\end{figure}
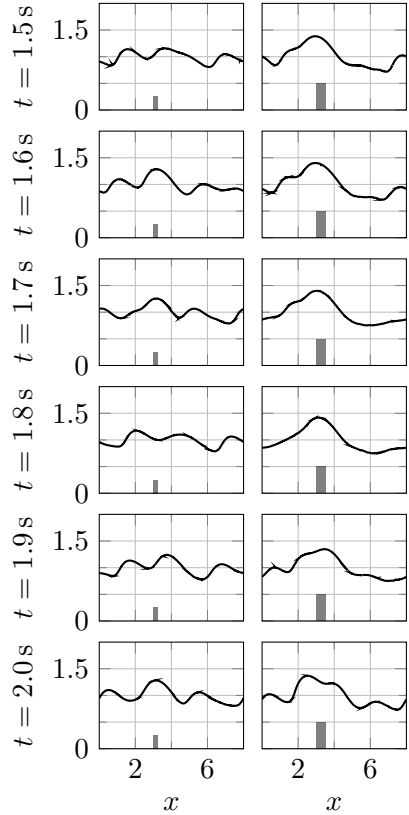

We observed, that all microstructures retained the film flow and greatly altered the film surface even for low Reynolds numbers.
However, this retaining before the microstructures led to a single, large hump or ridge already observed for example by~\cite{Baxter2009, Blyth2006,Tseluiko2013}.
Compared to the film on the smooth plate, all three rectangular structures inhibited the formation of waves.
This is apparent from the first row in~\cref{fig:res:film2d_fft} where all the amplitudes are almost zero except for a sharp peak obtained for the smooth surface.
The onset of the formation of waves was shifted to higher Reynolds numbers.
Furthermore, we observed, that the larger the structure compared to the film thickness, the stronger the inhibition (for example compare column 1 and 3 from~\cref{fig:res:film2d}).
Finally, if waves were formed for a critical $\Rey$, the waves were much more irregular than in the smooth case (compare for example line 5 and 6 in column 3 from~\cref{fig:res:film2d}).
The dynamics of the waves can be observed in~\cref{fig:res:film2d_time}.
Here, the interfaces are displayed for the small and medium rectangular structure ($h=0.25$ and $h=0.5$) at six instances in time between \SI{1.5}{\second} and \SI{2.0}{\second}.

For $h=0.25$, all films displayed very similar behavior for all $\Rey$, see the columns 2 and 5 in~\cref{fig:res:film2d}. 
The specific effect of a small triangular compared to a small rectangular microstructure is negligible.
For $\Rey<30$ only a slight retaining of the film could be observed and no waves, besides from the retaining of the film, occured. 
For $\Rey>40$ this picture suddenly changed and very irregular waves emerged for both structures. 
Interestingly, the structure of the waves did not really change with even higher $\Rey$, see the second and third column in~\cref{fig:res:film2d_fft}.

\subsection{Discussion}
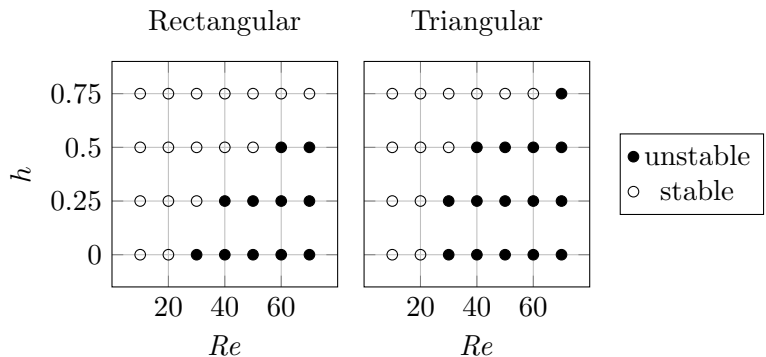
\begin{figure}
	\tikzsetnextfilename{stability}
	\centering
	\begin{tikzpicture}[scale=1.0]
		\begin{groupplot}[
			group style={
			    group size=2 by 1,
			    vertical sep=10pt,
				horizontal sep=10pt,
			    ylabels at=edge left,
				yticklabels at=edge left
			},
			width=3cm,
			height=3cm,
			scale only axis,
			grid=major,
			ylabel near ticks,
			xlabel near ticks
			,xmin=0, xmax=80, ymin=-0.15, ymax=0.9
			,xlabel=$\Rey$, ylabel=$h$
			,scatter/classes={%
					1={mark=*,black}, 0={mark=o,black}}
			,xtick={20,40,60}
			,ytick={0, 0.25,0.5,0.75}
			]%
			\nextgroupplot[title={Rectangular},legend to name=bla]
				\addplot[scatter,only marks, scatter src=explicit symbolic] table[x expr=\thisrow{Re}, y expr=\thisrow{struc}, meta=waves] {data/2d/stability_rec.dat};
				\addplot[scatter,only marks, scatter src=explicit symbolic] table[x expr=\thisrow{Re}, y expr=\thisrow{struc}, meta=waves] {data/2d/stability_smooth.dat};
				\legend{unstable, stable}
			\nextgroupplot[title={Triangular}]
						\addplot[scatter,only marks, scatter src=explicit symbolic] table[x expr=\thisrow{Re}, y expr=\thisrow{struc}, meta=waves] {data/2d/stability_tri.dat};
						\addplot[scatter,only marks, scatter src=explicit symbolic] table[x expr=\thisrow{Re}, y expr=\thisrow{struc}, meta=waves] {data/2d/stability_smooth.dat};
						\coordinate (top) at (rel axis cs:0,1);
						\coordinate (bot) at (rel axis cs:1,0);
		\end{groupplot}
		\path (top)--(bot) coordinate[midway] (group center);
    	\node[right=1em,inner sep=0pt] at(group center -| current bounding box.east) {\pgfplotslegendfromname{bla}};
	\end{tikzpicture}
		\caption{Instability diagrams for the rectangular (left) and the triangular structure, each for 3 different sizes and 7 Reynolds numbers. The results obtained for the smooth plate are added at $h=0$.}
	\label{fig:instability}
\end{figure}

The findings from~\cref{fig:res:film2d} and~\cref{fig:res:film2d_fft} are summarized in the instability diagrams in~\cref{fig:instability}.
Here, we plot the height of the microstructures against the Reynolds numbers.
The stable films are depicted by the circular markers, whereas the unstable films are marked as filled black dots.
For $h=0$ we show the results obtained from the smooth plate in both diagrams.

We clearly observe in~\cref{fig:res:film2d}, that, compared to the film on the smooth plate, all structures shift the formation of waves to higher $\Rey$.
Despite being small compared to the film height, the microstructures have a relative extreme effect on the liquid film.
They stabilize the film for smaller $\Rey$ but greatly destabilize the film for larger $\Rey$.
Furthermore, the inhibiting effect of larger structures compared to smaller structures is visible too.
In general, our data indicates, that the stabilizing effect, and the inhibition of waves, of the rectangular structures is slightly larger compared to the triangular structures.
Up to specific values of $h$ and $\Rey$ rectangular microstructures can prolong the formation of waves to higher $\Rey$ compared to the smooth plate (see rectangular, $h=0.5$, $\Rey<50$) or even completely suppress any waves and instabilities (see rectangular, $h=0.75$, all $\Rey$).
An explanation for this stabilizing effect is the greater dissipation in films overflowing sharp corners.
In case of the triangular and rectangular structure the liquid film has to overflow one two sharp corners, respectively.
The dissipated energy is missing from the film to form waves.
Furthermore, every structure must be bypassed by the liquid film and recirculation zones are formed before and after the structures.
All these factors might be contributing to the stabilization of the liquid film by microstructures with sharp corners.

\section{Conclusions}\label{sec:conclusion}
In the presented research, we were concerned with the stability of thin liquid films overflowing single microstructures with sharp corners.
The heights of the microstructures were comparable to the film thickness.
The dynamics of the two-phase flow were described by the coupling between the Cahn--Hilliard and the Navier--Stokes equations. 
The selected solution strategy guaranteed efficient and accurate simulations.
We validated our implementation against well-known experimental results.

We conducted simulations for Reynolds numbers $\Rey$ between 10 and 70.
The rectangular and triangular microstructures were of heights and widths of 0.25, 0.5, and 0.75 compared to the particular Nusselt film thickness.
In addition, simulations over a smooth surface were performed for comparison.
Our results show some very interesting stabilizing and destabilized effects.
Compared to the smooth plate, all structures shift the onset of waves to higher Reynolds numbers.
The specific effect of a small triangular compared to a small rectangular microstructure ($h=0.25$) is negligible.
Despite being small compared to the film height, the smaller microstructures greatly destabilized the film for higher $\Rey$.
Furthermore, the larger the structure compared to the film thickness, the stronger the effect of the inhibition of the formation of waves. 
Finally, if waves are formed for a critical $\Rey$, the waves seem to be much more irregular than in the smooth case.
In general, the inhibition of waves is stronger in the rectangular case compared to the triangular structure.
In these cases the microstructures act as stabilizer for the liquid film.

Our research indicates a strong influence of the size as well as the geometrical shape of the microstructures on the stability of the liquid film.
One stabilizing effect might be the dissipation of energy in the film while flowing over these sharp corners.
In case of the triangular and rectangular structure the liquid film has to overflow one and two sharp corners, respectively.
If could be valuable to investigate more complex shapes perhaps with additional corners.

\appendix
\section{Validation against~\citet{Wierschem2010}}\label{sec:validation}
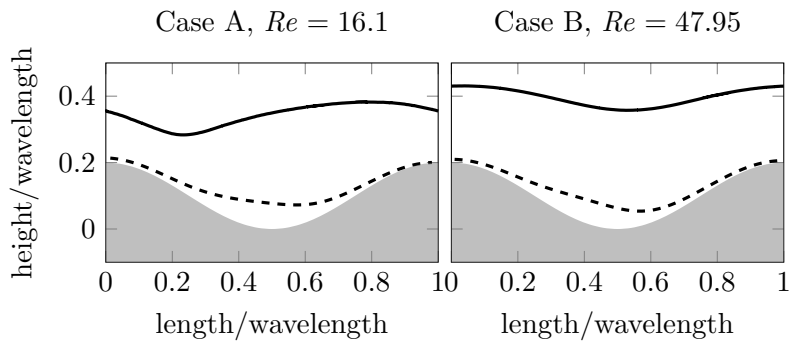
\begin{figure}[!h]
\tikzsetnextfilename{wierschem}
\centering
\begin{tikzpicture}
	\centering
	\begin{groupplot}[
			group style = {
				group size = 2 by 1
				,vertical sep=10pt,horizontal sep=5pt
				,ylabels at=edge left
               	,yticklabels at=edge left 
			},
    		width = 6cm
			,xlabel near ticks
			,ylabel near ticks
			,unit vector ratio=1 1 1
			, xmin=0, xmax=1
			, ymin=-0.1, ymax=0.5
			, xlabel={length/wavelength}
			, ylabel={height/wavelength}
			]
			\nextgroupplot[title={Case A, $\Rey=16.1$}]
				\addplot[dashed, very thick] table [x index = 2, y index = 3] {data/wierschem2010/wierschem2010_Re16.1_Cn0.04.dat};
				\addplot[very thick, smooth] plot file{data/wierschem2010/wierschem2010_Re16.1_Cn0.04_isoline_phi0.dat};
				\fill[opacity=0.5, gray, domain=0:1,smooth,variable=\x] (0, -0.1) -- plot ({\x},{0.1+0.1*cos(deg(2*pi*\x))}) -- (1,-0.1) -- cycle;
			\nextgroupplot[title={Case B, $\Rey=47.95$}]
				\addplot[dashed, very thick] table [x index = 2, y index = 3] {data/wierschem2010/wierschem2010_Re47.95_Cn0.04.dat};
				\addplot[very thick, smooth] plot file{data/wierschem2010/wierschem2010_Re47.95_Cn0.04_isoline_phi0.dat};
				\fill[opacity=0.5, gray, domain=0:1,smooth,variable=\x] (0, -0.1) -- plot ({\x},{0.1+0.1*cos(deg(2*pi*\x))}) -- (1,-0.1) -- cycle;
		\end{groupplot}
\end{tikzpicture}
\caption{Liquid film flowing over a deep sinusoidal corrugation. Corresponds to the experimental results in~\citet{Wierschem2010} Figure 3(b,d). The solid and dashed line represent the gas-liquid interface and the stream lines which separates the recirculation zone from the overflowing film.}	
\label{fig:res:wierschem}
\end{figure}

We validated our model and implementation by comparing our results with the experiments reported by~\cite{Wierschem2010}.
Following the recent work by~ \citet{Dietze2019} as well as the review by~\cite{Aksel2018} the used test case is well established.
In the experiments a film of silicone oil overflowed a deep sinusoidal corrugation.
The surface was inclined at \SI{8}{\degree} to the horizontal.
Similar to~\cite{Dietze2019} the Reynolds numbers $\Rey=16.1$ and $\Rey=47.95$ were simulated.
\Cref{tab:wierschemtable} lists the parameters used in the simulations.
Our simulation results are shown in~\cref{fig:res:wierschem}.
They correspond to the experimental results in~\citet{Wierschem2010} Figure 3(b,d).
The solid and dashed lines represent the gas-liquid interface and the stream line which separates the recirculation zone from the overflowing film.
Our results are very similar to both the experimental results by~\citet{Wierschem2010} and numerical results by~\citet{Dietze2019} (see Figure 14 therein).
The surface shape of the films including the positions of the minimum are accurately predicted.
In the corrugation through separation eddies are formed in similar sizes as in the experiment and the simulation.
\begin{table}[!h]
		\centering
\begin{tabular}{ccccccc}
\toprule
		Case&$\Rey$&$\Ca$&$\A$&$\Cn$&$\Pe_\epsilon$&$\delta_{init}$\\
\midrule
		A&16.1&0.063&0.99&0.04&147&0.00244\\
		B&47.95&0.130&0.99&0.04&304&0.00297\\
\bottomrule
\end{tabular}
\caption{Parameters used in the validation case. $\delta_{init}$ represents the initial film height taken from~\cite{Dietze2019}.}
\label{tab:wierschemtable}
\end{table}

\begin{acknowledgments}
The authors acknowledge the North-German Supercomputing Alliance (HLRN) for providing HPC resources that have contributed to the research results reported in this paper and thank the German Research Foundation (DFG) for the financial support within the project RE 1705/16-1.
Furthermore, we thank Christian Kahle for the fruitful discussions about the Cahn--Hilliard--Navier--Stokes equations.
\end{acknowledgments}

\bibliography{literature}

\begin{thebibliography}{45}%
\makeatletter
\providecommand \@ifxundefined [1]{%
 \@ifx{#1\undefined}
}%
\providecommand \@ifnum [1]{%
 \ifnum #1\expandafter \@firstoftwo
 \else \expandafter \@secondoftwo
 \fi
}%
\providecommand \@ifx [1]{%
 \ifx #1\expandafter \@firstoftwo
 \else \expandafter \@secondoftwo
 \fi
}%
\providecommand \natexlab [1]{#1}%
\providecommand \enquote  [1]{``#1''}%
\providecommand \bibnamefont  [1]{#1}%
\providecommand \bibfnamefont [1]{#1}%
\providecommand \citenamefont [1]{#1}%
\providecommand \href@noop [0]{\@secondoftwo}%
\providecommand \href [0]{\begingroup \@sanitize@url \@href}%
\providecommand \@href[1]{\@@startlink{#1}\@@href}%
\providecommand \@@href[1]{\endgroup#1\@@endlink}%
\providecommand \@sanitize@url [0]{\catcode `\\12\catcode `\$12\catcode
  `\&12\catcode `\#12\catcode `\^12\catcode `\_12\catcode `\%12\relax}%
\providecommand \@@startlink[1]{}%
\providecommand \@@endlink[0]{}%
\providecommand \url  [0]{\begingroup\@sanitize@url \@url }%
\providecommand \@url [1]{\endgroup\@href {#1}{\urlprefix }}%
\providecommand \urlprefix  [0]{URL }%
\providecommand \Eprint [0]{\href }%
\providecommand \doibase [0]{http://dx.doi.org/}%
\providecommand \selectlanguage [0]{\@gobble}%
\providecommand \bibinfo  [0]{\@secondoftwo}%
\providecommand \bibfield  [0]{\@secondoftwo}%
\providecommand \translation [1]{[#1]}%
\providecommand \BibitemOpen [0]{}%
\providecommand \bibitemStop [0]{}%
\providecommand \bibitemNoStop [0]{.\EOS\space}%
\providecommand \EOS [0]{\spacefactor3000\relax}%
\providecommand \BibitemShut  [1]{\csname bibitem#1\endcsname}%
\let\auto@bib@innerbib\@empty
\bibitem [{\citenamefont {Aksel}\ and\ \citenamefont
  {Sch{\"{o}}rner}(2018)}]{Aksel2018}%
  \BibitemOpen
  \bibfield  {author} {\bibinfo {author} {\bibfnamefont {N.}~\bibnamefont
  {Aksel}}\ and\ \bibinfo {author} {\bibfnamefont {M.}~\bibnamefont
  {Sch{\"{o}}rner}},\ }\href {\doibase 10.1007/s00707-018-2146-y} {\bibfield
  {journal} {\bibinfo  {journal} {Acta Mechanica}\ }\textbf {\bibinfo {volume}
  {229}},\ \bibinfo {pages} {1453} (\bibinfo {year} {2018})}\BibitemShut
  {NoStop}%
\bibitem [{\citenamefont {Blyth}\ and\ \citenamefont
  {Pozrikidis}(2006)}]{Blyth2006}%
  \BibitemOpen
  \bibfield  {author} {\bibinfo {author} {\bibfnamefont {M.~G.}\ \bibnamefont
  {Blyth}}\ and\ \bibinfo {author} {\bibfnamefont {C.}~\bibnamefont
  {Pozrikidis}},\ }\href {\doibase 10.1063/1.2198749} {\bibfield  {journal}
  {\bibinfo  {journal} {Physics of Fluids}\ }\textbf {\bibinfo {volume} {18}}
  (\bibinfo {year} {2006}),\ 10.1063/1.2198749}\BibitemShut {NoStop}%
\bibitem [{\citenamefont {Veremieiev}\ \emph {et~al.}(2015)\citenamefont
  {Veremieiev}, \citenamefont {Thompson},\ and\ \citenamefont
  {Gaskell}}]{Veremieiev2015}%
  \BibitemOpen
  \bibfield  {author} {\bibinfo {author} {\bibfnamefont {S.}~\bibnamefont
  {Veremieiev}}, \bibinfo {author} {\bibfnamefont {H.~M.}\ \bibnamefont
  {Thompson}}, \ and\ \bibinfo {author} {\bibfnamefont {P.~H.}\ \bibnamefont
  {Gaskell}},\ }\href {\doibase 10.1016/j.compfluid.2015.08.016} {\bibfield
  {journal} {\bibinfo  {journal} {Computers {\&} Fluids}\ }\textbf {\bibinfo
  {volume} {122}},\ \bibinfo {pages} {66} (\bibinfo {year} {2015})}\BibitemShut
  {NoStop}%
\bibitem [{\citenamefont {Craster}\ and\ \citenamefont
  {Matar}(2009)}]{Craster2009}%
  \BibitemOpen
  \bibfield  {author} {\bibinfo {author} {\bibfnamefont {R.~V.}\ \bibnamefont
  {Craster}}\ and\ \bibinfo {author} {\bibfnamefont {O.~K.}\ \bibnamefont
  {Matar}},\ }\href {\doibase 10.1103/RevModPhys.81.1131} {\bibfield  {journal}
  {\bibinfo  {journal} {Reviews of Modern Physics}\ }\textbf {\bibinfo {volume}
  {81}},\ \bibinfo {pages} {1131} (\bibinfo {year} {2009})}\BibitemShut
  {NoStop}%
\bibitem [{\citenamefont {Brunold}\ \emph {et~al.}(1989)\citenamefont
  {Brunold}, \citenamefont {Hunns}, \citenamefont {Mackley},\ and\
  \citenamefont {Thompson}}]{Brunold1989}%
  \BibitemOpen
  \bibfield  {author} {\bibinfo {author} {\bibfnamefont {C.}~\bibnamefont
  {Brunold}}, \bibinfo {author} {\bibfnamefont {J.}~\bibnamefont {Hunns}},
  \bibinfo {author} {\bibfnamefont {M.}~\bibnamefont {Mackley}}, \ and\
  \bibinfo {author} {\bibfnamefont {J.}~\bibnamefont {Thompson}},\ }\href
  {\doibase 10.1016/0009-2509(89)87022-8} {\bibfield  {journal} {\bibinfo
  {journal} {Chemical Engineering Science}\ }\textbf {\bibinfo {volume} {44}},\
  \bibinfo {pages} {1227} (\bibinfo {year} {1989})}\BibitemShut {NoStop}%
\bibitem [{\citenamefont {Ozgoren}(2006)}]{Ozgoren2006}%
  \BibitemOpen
  \bibfield  {author} {\bibinfo {author} {\bibfnamefont {M.}~\bibnamefont
  {Ozgoren}},\ }\href {\doibase 10.1016/j.flowmeasinst.2005.11.005} {\bibfield
  {journal} {\bibinfo  {journal} {Flow Measurement and Instrumentation}\
  }\textbf {\bibinfo {volume} {17}},\ \bibinfo {pages} {225} (\bibinfo {year}
  {2006})}\BibitemShut {NoStop}%
\bibitem [{\citenamefont {Hu}\ \emph {et~al.}(2006)\citenamefont {Hu},
  \citenamefont {Zhou},\ and\ \citenamefont {Dalton}}]{Hu2006}%
  \BibitemOpen
  \bibfield  {author} {\bibinfo {author} {\bibfnamefont {J.~C.}\ \bibnamefont
  {Hu}}, \bibinfo {author} {\bibfnamefont {Y.}~\bibnamefont {Zhou}}, \ and\
  \bibinfo {author} {\bibfnamefont {C.}~\bibnamefont {Dalton}},\ }\href
  {\doibase 10.1007/s00348-005-0052-2} {\bibfield  {journal} {\bibinfo
  {journal} {Experiments in Fluids}\ }\textbf {\bibinfo {volume} {40}},\
  \bibinfo {pages} {106} (\bibinfo {year} {2006})}\BibitemShut {NoStop}%
\bibitem [{\citenamefont {Alam}\ \emph {et~al.}(2011)\citenamefont {Alam},
  \citenamefont {Zhou},\ and\ \citenamefont {Wang}}]{Alam2011}%
  \BibitemOpen
  \bibfield  {author} {\bibinfo {author} {\bibfnamefont {M.~M.}\ \bibnamefont
  {Alam}}, \bibinfo {author} {\bibfnamefont {Y.}~\bibnamefont {Zhou}}, \ and\
  \bibinfo {author} {\bibfnamefont {X.~W.}\ \bibnamefont {Wang}},\ }\href
  {\doibase 10.1017/S0022112010005288} {\bibfield  {journal} {\bibinfo
  {journal} {Journal of Fluid Mechanics}\ }\textbf {\bibinfo {volume} {669}},\
  \bibinfo {pages} {432} (\bibinfo {year} {2011})}\BibitemShut {NoStop}%
\bibitem [{\citenamefont {Tseluiko}\ \emph {et~al.}(2013)\citenamefont
  {Tseluiko}, \citenamefont {Blyth},\ and\ \citenamefont
  {Papageorgiou}}]{Tseluiko2013}%
  \BibitemOpen
  \bibfield  {author} {\bibinfo {author} {\bibfnamefont {D.}~\bibnamefont
  {Tseluiko}}, \bibinfo {author} {\bibfnamefont {M.~G.}\ \bibnamefont {Blyth}},
  \ and\ \bibinfo {author} {\bibfnamefont {D.~T.}\ \bibnamefont
  {Papageorgiou}},\ }\href {\doibase 10.1017/jfm.2013.331} {\bibfield
  {journal} {\bibinfo  {journal} {Journal of Fluid Mechanics}\ }\textbf
  {\bibinfo {volume} {729}},\ \bibinfo {pages} {638} (\bibinfo {year}
  {2013})}\BibitemShut {NoStop}%
\bibitem [{\citenamefont {Gaskell}\ \emph {et~al.}(2004)\citenamefont
  {Gaskell}, \citenamefont {Jimack}, \citenamefont {Sellier}, \citenamefont
  {Thompson},\ and\ \citenamefont {Wilson}}]{Gaskell2004}%
  \BibitemOpen
  \bibfield  {author} {\bibinfo {author} {\bibfnamefont {P.~H.}\ \bibnamefont
  {Gaskell}}, \bibinfo {author} {\bibfnamefont {P.~K.}\ \bibnamefont {Jimack}},
  \bibinfo {author} {\bibfnamefont {M.}~\bibnamefont {Sellier}}, \bibinfo
  {author} {\bibfnamefont {H.~M.}\ \bibnamefont {Thompson}}, \ and\ \bibinfo
  {author} {\bibfnamefont {M.~C.~T.}\ \bibnamefont {Wilson}},\ }\href {\doibase
  10.1017/S0022112004009425} {\bibfield  {journal} {\bibinfo  {journal}
  {Journal of Fluid Mechanics}\ }\textbf {\bibinfo {volume} {509}},\ \bibinfo
  {pages} {253} (\bibinfo {year} {2004})}\BibitemShut {NoStop}%
\bibitem [{\citenamefont {Veremieiev}\ \emph {et~al.}(2010)\citenamefont
  {Veremieiev}, \citenamefont {Thompson}, \citenamefont {Lee},\ and\
  \citenamefont {Gaskell}}]{Veremieiev2010}%
  \BibitemOpen
  \bibfield  {author} {\bibinfo {author} {\bibfnamefont {S.}~\bibnamefont
  {Veremieiev}}, \bibinfo {author} {\bibfnamefont {H.~M.}\ \bibnamefont
  {Thompson}}, \bibinfo {author} {\bibfnamefont {Y.~C.}\ \bibnamefont {Lee}}, \
  and\ \bibinfo {author} {\bibfnamefont {P.~H.}\ \bibnamefont {Gaskell}},\
  }\href {\doibase 10.1016/j.compfluid.2009.09.007} {\bibfield  {journal}
  {\bibinfo  {journal} {Computers and Fluids}\ }\textbf {\bibinfo {volume}
  {39}},\ \bibinfo {pages} {431} (\bibinfo {year} {2010})}\BibitemShut
  {NoStop}%
\bibitem [{\citenamefont {Abels}\ \emph {et~al.}(2012)\citenamefont {Abels},
  \citenamefont {Garcke},\ and\ \citenamefont {Gr{\"u}n}}]{Abels2012}%
  \BibitemOpen
  \bibfield  {author} {\bibinfo {author} {\bibfnamefont {H.}~\bibnamefont
  {Abels}}, \bibinfo {author} {\bibfnamefont {H.}~\bibnamefont {Garcke}}, \
  and\ \bibinfo {author} {\bibfnamefont {G.}~\bibnamefont {Gr{\"u}n}},\ }\href
  {\doibase 10.1142/S0218202511500138} {\bibfield  {journal} {\bibinfo
  {journal} {Mathematical Models and Methods in Applied Sciences}\ }\textbf
  {\bibinfo {volume} {22}},\ \bibinfo {pages} {1150013(40)} (\bibinfo {year}
  {2012})}\BibitemShut {NoStop}%
\bibitem [{\citenamefont {Bonart}\ \emph
  {et~al.}(2019{\natexlab{a}})\citenamefont {Bonart}, \citenamefont {Kahle},\
  and\ \citenamefont {Repke}}]{Bonart2019b}%
  \BibitemOpen
  \bibfield  {author} {\bibinfo {author} {\bibfnamefont {H.}~\bibnamefont
  {Bonart}}, \bibinfo {author} {\bibfnamefont {C.}~\bibnamefont {Kahle}}, \
  and\ \bibinfo {author} {\bibfnamefont {J.-U.}\ \bibnamefont {Repke}},\ }\href
  {\doibase 10.1016/j.jcp.2019.108959} {\bibfield  {journal} {\bibinfo
  {journal} {Journal of Computational Physics}\ }\textbf {\bibinfo {volume}
  {399}},\ \bibinfo {pages} {108959} (\bibinfo {year} {2019}{\natexlab{a}})},\
  \Eprint {http://arxiv.org/abs/1809.06689} {arXiv:1809.06689} \BibitemShut
  {NoStop}%
\bibitem [{\citenamefont {Jacqmin}(1999)}]{Jacqmin1999}%
  \BibitemOpen
  \bibfield  {author} {\bibinfo {author} {\bibfnamefont {D.}~\bibnamefont
  {Jacqmin}},\ }\href {\doibase 10.1006/jcph.1999.6332} {\bibfield  {journal}
  {\bibinfo  {journal} {Journal of Computational Physics}\ }\textbf {\bibinfo
  {volume} {155}},\ \bibinfo {pages} {96} (\bibinfo {year} {1999})}\BibitemShut
  {NoStop}%
\bibitem [{\citenamefont {W{\"{o}}rner}(2012)}]{Worner2012}%
  \BibitemOpen
  \bibfield  {author} {\bibinfo {author} {\bibfnamefont {M.}~\bibnamefont
  {W{\"{o}}rner}},\ }\href {\doibase 10.1007/s10404-012-0940-8} {\bibfield
  {journal} {\bibinfo  {journal} {Microfluidics and Nanofluidics}\ }\textbf
  {\bibinfo {volume} {12}},\ \bibinfo {pages} {841} (\bibinfo {year}
  {2012})}\BibitemShut {NoStop}%
\bibitem [{\citenamefont {Anderson}\ \emph {et~al.}(1998)\citenamefont
  {Anderson}, \citenamefont {McFadden},\ and\ \citenamefont
  {Wheeler}}]{Anderson1998}%
  \BibitemOpen
  \bibfield  {author} {\bibinfo {author} {\bibfnamefont {D.~M.}\ \bibnamefont
  {Anderson}}, \bibinfo {author} {\bibfnamefont {G.~B.}\ \bibnamefont
  {McFadden}}, \ and\ \bibinfo {author} {\bibfnamefont {A.~A.}\ \bibnamefont
  {Wheeler}},\ }\href {\doibase 10.1146/annurev.fluid.30.1.139} {\bibfield
  {journal} {\bibinfo  {journal} {Annual Review of Fluid Mechanics}\ }\textbf
  {\bibinfo {volume} {30}},\ \bibinfo {pages} {139} (\bibinfo {year}
  {1998})}\BibitemShut {NoStop}%
\bibitem [{\citenamefont {He}\ and\ \citenamefont {Kasagi}(2008)}]{He2008}%
  \BibitemOpen
  \bibfield  {author} {\bibinfo {author} {\bibfnamefont {Q.}~\bibnamefont
  {He}}\ and\ \bibinfo {author} {\bibfnamefont {N.}~\bibnamefont {Kasagi}},\
  }\href {\doibase 10.1016/j.fluiddyn.2008.01.002} {\bibfield  {journal}
  {\bibinfo  {journal} {Fluid Dynamics Research}\ }\textbf {\bibinfo {volume}
  {40}},\ \bibinfo {pages} {497} (\bibinfo {year} {2008})}\BibitemShut
  {NoStop}%
\bibitem [{\citenamefont {Jamshidi}\ \emph {et~al.}(2018)\citenamefont
  {Jamshidi}, \citenamefont {Heimel}, \citenamefont {Hasert}, \citenamefont
  {Cai}, \citenamefont {Deutschmann}, \citenamefont {Marschall},\ and\
  \citenamefont {W{\"{o}}rner}}]{Jamshidi2018}%
  \BibitemOpen
  \bibfield  {author} {\bibinfo {author} {\bibfnamefont {F.}~\bibnamefont
  {Jamshidi}}, \bibinfo {author} {\bibfnamefont {H.}~\bibnamefont {Heimel}},
  \bibinfo {author} {\bibfnamefont {M.}~\bibnamefont {Hasert}}, \bibinfo
  {author} {\bibfnamefont {X.}~\bibnamefont {Cai}}, \bibinfo {author}
  {\bibfnamefont {O.}~\bibnamefont {Deutschmann}}, \bibinfo {author}
  {\bibfnamefont {H.}~\bibnamefont {Marschall}}, \ and\ \bibinfo {author}
  {\bibfnamefont {M.}~\bibnamefont {W{\"{o}}rner}},\ }\href {\doibase
  10.1016/j.cpc.2018.10.015} {\bibfield  {journal} {\bibinfo  {journal}
  {Computer Physics Communications}\ } (\bibinfo {year} {2018}),\
  10.1016/j.cpc.2018.10.015}\BibitemShut {NoStop}%
\bibitem [{Note1()}]{Note1}%
  \BibitemOpen
  \bibinfo {note} {For better readability we omit the $\protect \mathaccentV
  {hat}05E\protect \_$ above all scaled variables and operators from now on. If
  not otherwise noted, starting from~\protect \cref {eq:M:1_NS1} all variables
  are scaled and dimensionless.}\BibitemShut {Stop}%
\bibitem [{\citenamefont {Minjeaud}(2013)}]{Minjeaud2013}%
  \BibitemOpen
  \bibfield  {author} {\bibinfo {author} {\bibfnamefont {S.}~\bibnamefont
  {Minjeaud}},\ }\href {\doibase 10.1002/num.21721} {\bibfield  {journal}
  {\bibinfo  {journal} {Numerical Methods for Partial Differential Equations}\
  }\textbf {\bibinfo {volume} {29}},\ \bibinfo {pages} {584} (\bibinfo {year}
  {2013})}\BibitemShut {NoStop}%
\bibitem [{\citenamefont {Gr{\"{u}}n}\ \emph {et~al.}(2016)\citenamefont
  {Gr{\"{u}}n}, \citenamefont {Guill{\'{e}}n-Gonz{\'{a}}lez},\ and\
  \citenamefont {Metzger}}]{Gruen2016}%
  \BibitemOpen
  \bibfield  {author} {\bibinfo {author} {\bibfnamefont {G.}~\bibnamefont
  {Gr{\"{u}}n}}, \bibinfo {author} {\bibfnamefont {F.}~\bibnamefont
  {Guill{\'{e}}n-Gonz{\'{a}}lez}}, \ and\ \bibinfo {author} {\bibfnamefont
  {S.}~\bibnamefont {Metzger}},\ }\href {\doibase 10.4208/cicp.scpde14.39s}
  {\bibfield  {journal} {\bibinfo  {journal} {Communications in Computational
  Physics}\ }\textbf {\bibinfo {volume} {19}},\ \bibinfo {pages} {1473}
  (\bibinfo {year} {2016})}\BibitemShut {NoStop}%
\bibitem [{\citenamefont {Shen}\ \emph {et~al.}(2015)\citenamefont {Shen},
  \citenamefont {Yang},\ and\ \citenamefont
  {Yu}}]{2015-ShenYangYu-EnergyStableSchemesForCHMCL-Stabilization}%
  \BibitemOpen
  \bibfield  {author} {\bibinfo {author} {\bibfnamefont {J.}~\bibnamefont
  {Shen}}, \bibinfo {author} {\bibfnamefont {X.}~\bibnamefont {Yang}}, \ and\
  \bibinfo {author} {\bibfnamefont {H.}~\bibnamefont {Yu}},\ }\href {\doibase
  10.1016/j.jcp.2014.12.046} {\bibfield  {journal} {\bibinfo  {journal}
  {Journal of Computational Physics}\ }\textbf {\bibinfo {volume} {284}},\
  \bibinfo {pages} {617} (\bibinfo {year} {2015})}\BibitemShut {NoStop}%
\bibitem [{\citenamefont {Ferziger}\ and\ \citenamefont
  {Peric}(2008)}]{Ferziger2008}%
  \BibitemOpen
  \bibfield  {author} {\bibinfo {author} {\bibfnamefont {J.~H.}\ \bibnamefont
  {Ferziger}}\ and\ \bibinfo {author} {\bibfnamefont {M.}~\bibnamefont
  {Peric}},\ }\href {\doibase 10.1007/978-3-540-68228-8} {\emph {\bibinfo
  {title} {Book}}}\ (\bibinfo  {publisher} {Springer Berlin Heidelberg},\
  \bibinfo {address} {Berlin, Heidelberg},\ \bibinfo {year} {2008})\ p.\
  \bibinfo {pages} {509},\ \Eprint {http://arxiv.org/abs/arXiv:1011.1669v3}
  {arXiv:arXiv:1011.1669v3} \BibitemShut {NoStop}%
\bibitem [{\citenamefont {Aland}(2014)}]{Aland2014}%
  \BibitemOpen
  \bibfield  {author} {\bibinfo {author} {\bibfnamefont {S.}~\bibnamefont
  {Aland}},\ }\href {\doibase 10.1016/j.jcp.2013.12.055} {\bibfield  {journal}
  {\bibinfo  {journal} {Journal of Computational Physics}\ }\textbf {\bibinfo
  {volume} {262}},\ \bibinfo {pages} {58} (\bibinfo {year} {2014})},\ \Eprint
  {http://arxiv.org/abs/arXiv:1307.2127v1} {arXiv:arXiv:1307.2127v1}
  \BibitemShut {NoStop}%
\bibitem [{\citenamefont {Aln{\ae}s}\ \emph {et~al.}(2015)\citenamefont
  {Aln{\ae}s}, \citenamefont {Blechta}, \citenamefont {Hake}, \citenamefont
  {Johansson}, \citenamefont {Kehlet}, \citenamefont {Logg}, \citenamefont
  {Richardson}, \citenamefont {Ring}, \citenamefont {Rognes},\ and\
  \citenamefont {Wells}}]{AlnaesBlechta2015a}%
  \BibitemOpen
  \bibfield  {author} {\bibinfo {author} {\bibfnamefont {M.~S.}\ \bibnamefont
  {Aln{\ae}s}}, \bibinfo {author} {\bibfnamefont {J.}~\bibnamefont {Blechta}},
  \bibinfo {author} {\bibfnamefont {J.}~\bibnamefont {Hake}}, \bibinfo {author}
  {\bibfnamefont {A.}~\bibnamefont {Johansson}}, \bibinfo {author}
  {\bibfnamefont {B.}~\bibnamefont {Kehlet}}, \bibinfo {author} {\bibfnamefont
  {A.}~\bibnamefont {Logg}}, \bibinfo {author} {\bibfnamefont {C.}~\bibnamefont
  {Richardson}}, \bibinfo {author} {\bibfnamefont {J.}~\bibnamefont {Ring}},
  \bibinfo {author} {\bibfnamefont {M.~E.}\ \bibnamefont {Rognes}}, \ and\
  \bibinfo {author} {\bibfnamefont {G.~N.}\ \bibnamefont {Wells}},\ }\href
  {\doibase 10.11588/ans.2015.100.20553} {\bibfield  {journal} {\bibinfo
  {journal} {Archive of Numerical Software}\ }\textbf {\bibinfo {volume} {3}}
  (\bibinfo {year} {2015}),\ 10.11588/ans.2015.100.20553}\BibitemShut {NoStop}%
\bibitem [{\citenamefont {Logg}\ \emph {et~al.}(2012)\citenamefont {Logg},
  \citenamefont {Mardal},\ and\ \citenamefont {Wells}}]{fenics_book}%
  \BibitemOpen
  \bibinfo {editor} {\bibfnamefont {A.}~\bibnamefont {Logg}}, \bibinfo {editor}
  {\bibfnamefont {K.-A.}\ \bibnamefont {Mardal}}, \ and\ \bibinfo {editor}
  {\bibfnamefont {G.}~\bibnamefont {Wells}},\ eds.,\ \href {\doibase
  10.1007/978-3-642-23099-8} {\emph {\bibinfo {title} {{Automated Solution of
  Differential Equations by the Finite Element Method - The FEniCS Book}}}},\
  \bibinfo {series} {Lecture Notes in Computational Science and Engineering},
  Vol.~\bibinfo {volume} {84}\ (\bibinfo  {publisher} {Springer},\ \bibinfo
  {year} {2012})\BibitemShut {NoStop}%
\bibitem [{\citenamefont {Balay}\ \emph
  {et~al.}(2018{\natexlab{a}})\citenamefont {Balay}, \citenamefont {Abhyankar},
  \citenamefont {Adams}, \citenamefont {Brown}, \citenamefont {Brune},
  \citenamefont {Buschelman}, \citenamefont {Dalcin}, \citenamefont {Eijkhout},
  \citenamefont {Gropp}, \citenamefont {Kaushik}, \citenamefont {Knepley},
  \citenamefont {May}, \citenamefont {McInnes}, \citenamefont {Mills},
  \citenamefont {Munson}, \citenamefont {Rupp}, \citenamefont {Sanan},
  \citenamefont {Smith}, \citenamefont {Zampini}, \citenamefont {Zhang},\ and\
  \citenamefont {Zhang}}]{petsc_webpage}%
  \BibitemOpen
  \bibfield  {author} {\bibinfo {author} {\bibfnamefont {S.}~\bibnamefont
  {Balay}}, \bibinfo {author} {\bibfnamefont {S.}~\bibnamefont {Abhyankar}},
  \bibinfo {author} {\bibfnamefont {M.}~\bibnamefont {Adams}}, \bibinfo
  {author} {\bibfnamefont {J.}~\bibnamefont {Brown}}, \bibinfo {author}
  {\bibfnamefont {P.}~\bibnamefont {Brune}}, \bibinfo {author} {\bibfnamefont
  {K.}~\bibnamefont {Buschelman}}, \bibinfo {author} {\bibfnamefont
  {L.}~\bibnamefont {Dalcin}}, \bibinfo {author} {\bibfnamefont
  {V.}~\bibnamefont {Eijkhout}}, \bibinfo {author} {\bibfnamefont
  {W.}~\bibnamefont {Gropp}}, \bibinfo {author} {\bibfnamefont
  {D.}~\bibnamefont {Kaushik}}, \bibinfo {author} {\bibfnamefont
  {M.}~\bibnamefont {Knepley}}, \bibinfo {author} {\bibfnamefont
  {D.}~\bibnamefont {May}}, \bibinfo {author} {\bibfnamefont {L.~C.}\
  \bibnamefont {McInnes}}, \bibinfo {author} {\bibfnamefont {R.~T.}\
  \bibnamefont {Mills}}, \bibinfo {author} {\bibfnamefont {T.}~\bibnamefont
  {Munson}}, \bibinfo {author} {\bibfnamefont {K.}~\bibnamefont {Rupp}},
  \bibinfo {author} {\bibfnamefont {P.}~\bibnamefont {Sanan}}, \bibinfo
  {author} {\bibfnamefont {B.}~\bibnamefont {Smith}}, \bibinfo {author}
  {\bibfnamefont {S.}~\bibnamefont {Zampini}}, \bibinfo {author} {\bibfnamefont
  {H.}~\bibnamefont {Zhang}}, \ and\ \bibinfo {author} {\bibfnamefont
  {H.}~\bibnamefont {Zhang}},\ }\href {http://www.mcs.anl.gov/petsc} {\enquote
  {\bibinfo {title} {{PETSc Web page}},}\ }\bibinfo {howpublished}
  {{http://www.mcs.anl.gov/petsc}} (\bibinfo {year}
  {2018}{\natexlab{a}})\BibitemShut {NoStop}%
\bibitem [{\citenamefont {Balay}\ \emph
  {et~al.}(2018{\natexlab{b}})\citenamefont {Balay}, \citenamefont {Abhyankar},
  \citenamefont {Adams}, \citenamefont {Brown}, \citenamefont {Brune},
  \citenamefont {Buschelman}, \citenamefont {Dalcin}, \citenamefont {Eijkhout},
  \citenamefont {Gropp}, \citenamefont {Kaushik}, \citenamefont {Knepley},
  \citenamefont {May}, \citenamefont {McInnes}, \citenamefont {Mills},
  \citenamefont {Munson}, \citenamefont {Rupp}, \citenamefont {Sanan},
  \citenamefont {Smith}, \citenamefont {Zampini}, \citenamefont {Zhang},\ and\
  \citenamefont {Zhang}}]{petsc-user-ref}%
  \BibitemOpen
  \bibfield  {author} {\bibinfo {author} {\bibfnamefont {S.}~\bibnamefont
  {Balay}}, \bibinfo {author} {\bibfnamefont {S.}~\bibnamefont {Abhyankar}},
  \bibinfo {author} {\bibfnamefont {M.}~\bibnamefont {Adams}}, \bibinfo
  {author} {\bibfnamefont {J.}~\bibnamefont {Brown}}, \bibinfo {author}
  {\bibfnamefont {P.}~\bibnamefont {Brune}}, \bibinfo {author} {\bibfnamefont
  {K.}~\bibnamefont {Buschelman}}, \bibinfo {author} {\bibfnamefont
  {L.}~\bibnamefont {Dalcin}}, \bibinfo {author} {\bibfnamefont
  {V.}~\bibnamefont {Eijkhout}}, \bibinfo {author} {\bibfnamefont
  {W.}~\bibnamefont {Gropp}}, \bibinfo {author} {\bibfnamefont
  {D.}~\bibnamefont {Kaushik}}, \bibinfo {author} {\bibfnamefont
  {M.}~\bibnamefont {Knepley}}, \bibinfo {author} {\bibfnamefont
  {D.}~\bibnamefont {May}}, \bibinfo {author} {\bibfnamefont {L.~C.}\
  \bibnamefont {McInnes}}, \bibinfo {author} {\bibfnamefont {R.~T.}\
  \bibnamefont {Mills}}, \bibinfo {author} {\bibfnamefont {T.}~\bibnamefont
  {Munson}}, \bibinfo {author} {\bibfnamefont {K.}~\bibnamefont {Rupp}},
  \bibinfo {author} {\bibfnamefont {P.}~\bibnamefont {Sanan}}, \bibinfo
  {author} {\bibfnamefont {B.}~\bibnamefont {Smith}}, \bibinfo {author}
  {\bibfnamefont {S.}~\bibnamefont {Zampini}}, \bibinfo {author} {\bibfnamefont
  {H.}~\bibnamefont {Zhang}}, \ and\ \bibinfo {author} {\bibfnamefont
  {H.}~\bibnamefont {Zhang}},\ }\href {http://www.mcs.anl.gov/petsc} {\emph
  {\bibinfo {title} {{{\{}PETS{\}}c Users Manual}}}},\ \bibinfo {type} {Tech.
  Rep.}\ \bibinfo {number} {ANL-95/11 - Revision 3.9}\ (\bibinfo  {institution}
  {Argonne National Laboratory},\ \bibinfo {year} {2018})\BibitemShut {NoStop}%
\bibitem [{\citenamefont {Balay}\ \emph {et~al.}(1997)\citenamefont {Balay},
  \citenamefont {Gropp}, \citenamefont {McInnes},\ and\ \citenamefont
  {Smith}}]{petsc-efficient}%
  \BibitemOpen
  \bibfield  {author} {\bibinfo {author} {\bibfnamefont {S.}~\bibnamefont
  {Balay}}, \bibinfo {author} {\bibfnamefont {W.~D.}\ \bibnamefont {Gropp}},
  \bibinfo {author} {\bibfnamefont {L.~C.}\ \bibnamefont {McInnes}}, \ and\
  \bibinfo {author} {\bibfnamefont {B.~F.}\ \bibnamefont {Smith}},\ }in\
  \href@noop {} {\emph {\bibinfo {booktitle} {Modern Software Tools in
  Scientific Computing}}},\ \bibinfo {editor} {edited by\ \bibinfo {editor}
  {\bibfnamefont {E.}~\bibnamefont {Arge}}, \bibinfo {editor} {\bibfnamefont
  {A.~M.}\ \bibnamefont {Bruaset}}, \ and\ \bibinfo {editor} {\bibfnamefont
  {H.~P.}\ \bibnamefont {Langtangen}}}\ (\bibinfo  {publisher}
  {Birkh{\"{a}}user Press},\ \bibinfo {year} {1997})\ pp.\ \bibinfo {pages}
  {163--202}\BibitemShut {NoStop}%
\bibitem [{\citenamefont {Amestoy}\ \emph {et~al.}(2001)\citenamefont
  {Amestoy}, \citenamefont {Duff}, \citenamefont {Koster},\ and\ \citenamefont
  {L'Excellent}}]{mumps_1}%
  \BibitemOpen
  \bibfield  {author} {\bibinfo {author} {\bibfnamefont {P.~R.}\ \bibnamefont
  {Amestoy}}, \bibinfo {author} {\bibfnamefont {I.~S.}\ \bibnamefont {Duff}},
  \bibinfo {author} {\bibfnamefont {J.}~\bibnamefont {Koster}}, \ and\ \bibinfo
  {author} {\bibfnamefont {J.-Y.}\ \bibnamefont {L'Excellent}},\ }\href@noop {}
  {\bibfield  {journal} {\bibinfo  {journal} {SIAM Journal on Matrix Analysis
  and Applications}\ }\textbf {\bibinfo {volume} {23}},\ \bibinfo {pages} {15}
  (\bibinfo {year} {2001})}\BibitemShut {NoStop}%
\bibitem [{\citenamefont {Amestoy}\ \emph {et~al.}(2006)\citenamefont
  {Amestoy}, \citenamefont {Guermouche}, \citenamefont {L'Excellent},\ and\
  \citenamefont {Pralet}}]{mumps_2}%
  \BibitemOpen
  \bibfield  {author} {\bibinfo {author} {\bibfnamefont {P.~R.}\ \bibnamefont
  {Amestoy}}, \bibinfo {author} {\bibfnamefont {A.}~\bibnamefont {Guermouche}},
  \bibinfo {author} {\bibfnamefont {J.-Y.}\ \bibnamefont {L'Excellent}}, \ and\
  \bibinfo {author} {\bibfnamefont {S.}~\bibnamefont {Pralet}},\ }\href@noop {}
  {\bibfield  {journal} {\bibinfo  {journal} {Parallel Computing}\ }\textbf
  {\bibinfo {volume} {32}},\ \bibinfo {pages} {136} (\bibinfo {year}
  {2006})}\BibitemShut {NoStop}%
\bibitem [{\citenamefont {Boyanova}\ \emph {et~al.}(2012)\citenamefont
  {Boyanova}, \citenamefont {Do-Quang},\ and\ \citenamefont
  {Neytcheva}}]{Boyanova2012}%
  \BibitemOpen
  \bibfield  {author} {\bibinfo {author} {\bibfnamefont {P.}~\bibnamefont
  {Boyanova}}, \bibinfo {author} {\bibfnamefont {M.}~\bibnamefont {Do-Quang}},
  \ and\ \bibinfo {author} {\bibfnamefont {M.}~\bibnamefont {Neytcheva}},\
  }\href {\doibase 10.2478/cmam-2012-0001} {\bibfield  {journal} {\bibinfo
  {journal} {Computational Methods in Applied Mathematics}\ }\textbf {\bibinfo
  {volume} {12}},\ \bibinfo {pages} {1} (\bibinfo {year} {2012})}\BibitemShut
  {NoStop}%
\bibitem [{\citenamefont {Bosch}\ \emph {et~al.}(2018)\citenamefont {Bosch},
  \citenamefont {Kahle},\ and\ \citenamefont {Stoll}}]{Bosch2018}%
  \BibitemOpen
  \bibfield  {author} {\bibinfo {author} {\bibfnamefont {J.}~\bibnamefont
  {Bosch}}, \bibinfo {author} {\bibfnamefont {C.}~\bibnamefont {Kahle}}, \ and\
  \bibinfo {author} {\bibfnamefont {M.}~\bibnamefont {Stoll}},\ }\href
  {\doibase 10.4208/cicp.OA-2017-0037} {\bibfield  {journal} {\bibinfo
  {journal} {Communications in Computational Physics}\ }\textbf {\bibinfo
  {volume} {23}} (\bibinfo {year} {2018}),\ 10.4208/cicp.OA-2017-0037},\
  \Eprint {http://arxiv.org/abs/1610.03991} {arXiv:1610.03991} \BibitemShut
  {NoStop}%
\bibitem [{\citenamefont {Blechta}(2019)}]{Blechta2019}%
  \BibitemOpen
  \bibfield  {author} {\bibinfo {author} {\bibfnamefont {J.}~\bibnamefont
  {Blechta}},\ }\href@noop {} {\enquote {\bibinfo {title} {{Towards efficient
  numerical computation of flows of non-Newtonian fluids}},}\ } (\bibinfo
  {year} {2019})\BibitemShut {NoStop}%
\bibitem [{\citenamefont {Olshanskii}\ and\ \citenamefont
  {Vassilevski}(2007)}]{Olshanskii2007}%
  \BibitemOpen
  \bibfield  {author} {\bibinfo {author} {\bibfnamefont {M.~A.}\ \bibnamefont
  {Olshanskii}}\ and\ \bibinfo {author} {\bibfnamefont {Y.~V.}\ \bibnamefont
  {Vassilevski}},\ }\href {\doibase 10.1137/070679776} {\bibfield  {journal}
  {\bibinfo  {journal} {SIAM Journal on Scientific Computing}\ }\textbf
  {\bibinfo {volume} {29}},\ \bibinfo {pages} {2686} (\bibinfo {year}
  {2007})}\BibitemShut {NoStop}%
\bibitem [{\citenamefont {Elman}\ \emph {et~al.}(2008)\citenamefont {Elman},
  \citenamefont {Howle}, \citenamefont {Shadid}, \citenamefont {Shuttleworth},\
  and\ \citenamefont {Tuminaro}}]{Elman2008}%
  \BibitemOpen
  \bibfield  {author} {\bibinfo {author} {\bibfnamefont {H.}~\bibnamefont
  {Elman}}, \bibinfo {author} {\bibfnamefont {V.~E.}\ \bibnamefont {Howle}},
  \bibinfo {author} {\bibfnamefont {J.}~\bibnamefont {Shadid}}, \bibinfo
  {author} {\bibfnamefont {R.}~\bibnamefont {Shuttleworth}}, \ and\ \bibinfo
  {author} {\bibfnamefont {R.}~\bibnamefont {Tuminaro}},\ }\href {\doibase
  10.1016/j.jcp.2007.09.026} {\bibfield  {journal} {\bibinfo  {journal}
  {Journal of Computational Physics}\ }\textbf {\bibinfo {volume} {227}},\
  \bibinfo {pages} {1790} (\bibinfo {year} {2008})}\BibitemShut {NoStop}%
\bibitem [{\citenamefont {Bootland}\ \emph {et~al.}(2019)\citenamefont
  {Bootland}, \citenamefont {Bentley}, \citenamefont {Kees},\ and\
  \citenamefont {Wathen}}]{Bootland2019}%
  \BibitemOpen
  \bibfield  {author} {\bibinfo {author} {\bibfnamefont {N.}~\bibnamefont
  {Bootland}}, \bibinfo {author} {\bibfnamefont {A.}~\bibnamefont {Bentley}},
  \bibinfo {author} {\bibfnamefont {C.}~\bibnamefont {Kees}}, \ and\ \bibinfo
  {author} {\bibfnamefont {A.}~\bibnamefont {Wathen}},\ }\href {\doibase
  10.1137/17M1153674} {\bibfield  {journal} {\bibinfo  {journal} {SIAM Journal
  on Scientific Computing}\ }\textbf {\bibinfo {volume} {41}},\ \bibinfo
  {pages} {B843} (\bibinfo {year} {2019})},\ \Eprint
  {http://arxiv.org/abs/1710.08779} {arXiv:1710.08779} \BibitemShut {NoStop}%
\bibitem [{\citenamefont {Hysing}\ \emph {et~al.}(2009)\citenamefont {Hysing},
  \citenamefont {Turek}, \citenamefont {Kuzmin}, \citenamefont {Parolini},
  \citenamefont {Burman}, \citenamefont {Ganesan},\ and\ \citenamefont
  {Tobiska}}]{Hysing2009}%
  \BibitemOpen
  \bibfield  {author} {\bibinfo {author} {\bibfnamefont {S.}~\bibnamefont
  {Hysing}}, \bibinfo {author} {\bibfnamefont {S.}~\bibnamefont {Turek}},
  \bibinfo {author} {\bibfnamefont {D.}~\bibnamefont {Kuzmin}}, \bibinfo
  {author} {\bibfnamefont {N.}~\bibnamefont {Parolini}}, \bibinfo {author}
  {\bibfnamefont {E.}~\bibnamefont {Burman}}, \bibinfo {author} {\bibfnamefont
  {S.}~\bibnamefont {Ganesan}}, \ and\ \bibinfo {author} {\bibfnamefont
  {L.}~\bibnamefont {Tobiska}},\ }\href {\doibase 10.1002/fld.1934} {\bibfield
  {journal} {\bibinfo  {journal} {International Journal for Numerical Methods
  in Fluids}\ }\textbf {\bibinfo {volume} {60}},\ \bibinfo {pages} {1259}
  (\bibinfo {year} {2009})}\BibitemShut {NoStop}%
\bibitem [{\citenamefont {Bonart}\ \emph
  {et~al.}(2019{\natexlab{b}})\citenamefont {Bonart}, \citenamefont {Jung},
  \citenamefont {Kahle},\ and\ \citenamefont {Repke}}]{Bonart2019a}%
  \BibitemOpen
  \bibfield  {author} {\bibinfo {author} {\bibfnamefont {H.}~\bibnamefont
  {Bonart}}, \bibinfo {author} {\bibfnamefont {J.}~\bibnamefont {Jung}},
  \bibinfo {author} {\bibfnamefont {C.}~\bibnamefont {Kahle}}, \ and\ \bibinfo
  {author} {\bibfnamefont {J.-U.}\ \bibnamefont {Repke}},\ }\href {\doibase
  10.1002/ceat.201900029} {\bibfield  {journal} {\bibinfo  {journal} {Chemical
  Engineering {\&} Technology}\ }\textbf {\bibinfo {volume} {42}},\ \bibinfo
  {pages} {1381} (\bibinfo {year} {2019}{\natexlab{b}})}\BibitemShut {NoStop}%
\bibitem [{\citenamefont {Bonart}\ and\ \citenamefont
  {Repke}(2018)}]{Bonart2018a}%
  \BibitemOpen
  \bibfield  {author} {\bibinfo {author} {\bibfnamefont {H.}~\bibnamefont
  {Bonart}}\ and\ \bibinfo {author} {\bibfnamefont {J.-U.}\ \bibnamefont
  {Repke}},\ }\href {\doibase 10.3303/CET1869011} {\bibfield  {journal}
  {\bibinfo  {journal} {Chemical Engineering Transactions}\ }\textbf {\bibinfo
  {volume} {69}},\ \bibinfo {pages} {61} (\bibinfo {year} {2018})}\BibitemShut
  {NoStop}%
\bibitem [{\citenamefont {Wierschem}\ \emph {et~al.}(2010)\citenamefont
  {Wierschem}, \citenamefont {Pollak}, \citenamefont {Heining},\ and\
  \citenamefont {Aksel}}]{Wierschem2010}%
  \BibitemOpen
  \bibfield  {author} {\bibinfo {author} {\bibfnamefont {A.}~\bibnamefont
  {Wierschem}}, \bibinfo {author} {\bibfnamefont {T.}~\bibnamefont {Pollak}},
  \bibinfo {author} {\bibfnamefont {C.}~\bibnamefont {Heining}}, \ and\
  \bibinfo {author} {\bibfnamefont {N.}~\bibnamefont {Aksel}},\ }\href
  {\doibase 10.1063/1.3504374} {\bibfield  {journal} {\bibinfo  {journal}
  {Physics of Fluids}\ }\textbf {\bibinfo {volume} {22}},\ \bibinfo {pages}
  {113603} (\bibinfo {year} {2010})}\BibitemShut {NoStop}%
\bibitem [{\citenamefont {Dietze}(2019)}]{Dietze2019}%
  \BibitemOpen
  \bibfield  {author} {\bibinfo {author} {\bibfnamefont {G.~F.}\ \bibnamefont
  {Dietze}},\ }\href {\doibase 10.1017/jfm.2018.851} {\bibfield  {journal}
  {\bibinfo  {journal} {Journal of Fluid Mechanics}\ }\textbf {\bibinfo
  {volume} {859}},\ \bibinfo {pages} {1098} (\bibinfo {year}
  {2019})}\BibitemShut {NoStop}%
\bibitem [{\citenamefont {Geuzaine}\ and\ \citenamefont
  {Remacle}(2009)}]{Geuzaine2009}%
  \BibitemOpen
  \bibfield  {author} {\bibinfo {author} {\bibfnamefont {C.}~\bibnamefont
  {Geuzaine}}\ and\ \bibinfo {author} {\bibfnamefont {J.-F.}\ \bibnamefont
  {Remacle}},\ }\href {\doibase 10.1002/nme.2579} {\bibfield  {journal}
  {\bibinfo  {journal} {International Journal for Numerical Methods in
  Engineering}\ }\textbf {\bibinfo {volume} {79}},\ \bibinfo {pages} {1309}
  (\bibinfo {year} {2009})}\BibitemShut {NoStop}%
\bibitem [{\citenamefont {Cai}\ \emph {et~al.}(2015)\citenamefont {Cai},
  \citenamefont {Marschall}, \citenamefont {W{\"{o}}rner},\ and\ \citenamefont
  {Deutschmann}}]{Cai2015}%
  \BibitemOpen
  \bibfield  {author} {\bibinfo {author} {\bibfnamefont {X.}~\bibnamefont
  {Cai}}, \bibinfo {author} {\bibfnamefont {H.}~\bibnamefont {Marschall}},
  \bibinfo {author} {\bibfnamefont {M.}~\bibnamefont {W{\"{o}}rner}}, \ and\
  \bibinfo {author} {\bibfnamefont {O.}~\bibnamefont {Deutschmann}},\ }\href
  {\doibase 10.1002/ceat.201500089} {\bibfield  {journal} {\bibinfo  {journal}
  {Chemical Engineering {\&} Technology}\ }\textbf {\bibinfo {volume} {38}},\
  \bibinfo {pages} {1985} (\bibinfo {year} {2015})}\BibitemShut {NoStop}%
\bibitem [{\citenamefont {Baxter}\ \emph {et~al.}(2009)\citenamefont {Baxter},
  \citenamefont {Power}, \citenamefont {Cliffe},\ and\ \citenamefont
  {Hibberd}}]{Baxter2009}%
  \BibitemOpen
  \bibfield  {author} {\bibinfo {author} {\bibfnamefont {S.~J.}\ \bibnamefont
  {Baxter}}, \bibinfo {author} {\bibfnamefont {H.}~\bibnamefont {Power}},
  \bibinfo {author} {\bibfnamefont {K.~A.}\ \bibnamefont {Cliffe}}, \ and\
  \bibinfo {author} {\bibfnamefont {S.}~\bibnamefont {Hibberd}},\ }\href
  {\doibase 10.1063/1.3082218} {\bibfield  {journal} {\bibinfo  {journal}
  {Physics of Fluids}\ }\textbf {\bibinfo {volume} {21}} (\bibinfo {year}
  {2009}),\ 10.1063/1.3082218}\BibitemShut {NoStop}%
\end{thebibliography}%

\end{document}